\documentclass[usenatbib,useAMS]{mn2e}

\usepackage[dvips]{graphicx}
\usepackage{dcolumn}
\usepackage{times}

\title{Radio, optical and infrared observations of CLASS B0128+437}
\author[A.~D.~Biggs et al.]{A.~D.~Biggs,$^{1,2}$\thanks{E-mail:
    biggs@jive.nl} I.~W.~A. Browne,$^2$ N.~J.~Jackson,$^2$ T.~York,$^2$
    M.~A.~Norbury,$^2$
\newauthor J.~P.~McKean$^2$ and P.~M.~Phillips$^2$\\
$^1$Joint Institute for VLBI in Europe, Postbus 2, 7990 AA Dwingeloo,
The Netherlands\\
$^2$University of Manchester, Jodrell Bank Observatory, Macclesfield, Cheshire 
SK11 9DL}
\begin{document}
\maketitle
\begin{abstract}
We present new observations of the gravitational lens system CLASS
B0128+437 made in the optical, infrared and radio regimes. $Hubble
Space Telescope$ observations detect only a very faint, extended object
in $I$-band with no obvious emission from the lensed images visible; no
detection at all is made in $V$-band. The lens system is detected with
much higher signal to noise with UKIRT in $K$-band and, although
resolved, the resolution is not sufficient to allow the lensed images
and the lens galaxy to be separated. A careful astrometric calibration,
however, suggests that the peak of the infrared emission corresponds to
the two merging images A and B and therefore that the lensed images
dominate at infrared wavelengths. The new radio data consist of high
resolution VLBI radio images at three frequencies, 2.3, 5 and 8.4~GHz, 
made with the VLBA and the 100-m Effelsberg telescope. These reveal
that the lensed source consists of three well-defined
sub-components that are embedded in a more extended jet. Due to the
fact that the sub-components have different spectral indices it is
possible to determine, unambiguously, which part of each image
corresponds to the same source sub-component. Our main finding is that
one of the images, B, looks very different to the others, there being
no obvious division into separate sub-components and the image being
apparently both broader and smoother. This is a consequence we believe
of scatter-broadening in the ISM of the lensing galaxy. The large
number of multiply-imaged source sub-components also provide an
abundance of modelling constraints and we have attempted to fit an
SIE+external shear model to the data, as well as utilising the novel
method of \citeauthor{evans03}. It proves difficult in both cases,
however, to obtain a satisfactory fit which strongly suggests the
presence of sub-structure in the mass distribution of the lensing
galaxy, perhaps of the kind that is predicted by CDM theories of
structure formation.

\end{abstract}


\section{Introduction}

The Cosmic Lens All Sky Survey (CLASS), a radio survey of the
northern sky, has been exceptionally successful in discovering new
arcsec-scale gravitational lens systems, with a final tally of 22
\citep{myers03,browne03}. These systems can, using the method of
\citet{refsdal64}, be used individually to determine the Hubble
parameter \citep[e.g][]{kochanek03,burud02,treu02,fassnacht02} and
together to place constraints on the cosmological parameters $\Omega_0$
and $\lambda_0$ \citep[e.g][]{helbig99,chae02}. As well as this,
gravitational lenses are used to explore various aspects of
high-redshift galaxies i.e. the lens galaxies themselves
\citep[e.g.][]{kochanek00,lehar00}, including their distribution of
ionised and magnetised plasma, their radial mass profiles 
\citep[e.g][]{koopmans03b,winn03a,biggs03,rusin02,rusin01,munoz01} and 
the presence of massive substructures such as those predicted by CDM
theories of large-scale structure formation
\citep[e.g][]{mao98,bradac02,metcalf01,metcalf02a,metcalf02b,dalal02}.

B0128+437 \citep{phillips00} was discovered by CLASS and consists of
four images with a maximum separation of 540~mas
(Fig.~\ref{optical}). Initial modelling, in the absence of any
information regarding the lensing galaxy, demonstrated that a singular 
isothermal ellipsoidal mass distribution plus a large external shear
produces a good fit to the observed image positions and flux
ratios. Subsequent optical spectroscopy with the W.~M.~Keck telescope
\citep{mckean04} measured a redshift for the lensed
source of $z_s=3.12$ and identified a single line at a different
redshift that is presumably that of the lensing galaxy. The most likely
identification of this line (\mbox{[O\,{\sc ii}]}) gives a lens redshift
of $z_d=1.145$ although additional observations will be required to
confirm this. The radio spectrum of the lensed source has a pronounced
peak at around 1~GHz, thus making it a member of the Giga-hertz Peaked
Spectrum (GPS) class.

In this paper we present new optical, infrared and radio VLBI
observations of B0128+437 which demonstrate that this system is far
more interesting than the original 5-GHz MERLIN data suggested.
In the next section we show images made using the $Hubble Space
Telescope$ ($HST$) and the United Kingdom Infrared Telescope
(UKIRT). We then go on to describe new multi-frequency observations
(2.3, 5 and 8.4~GHz) made with the VLBA together with the 100-m
Effelsberg telescope and present the resulting maps of each of the
lensed images. The remainder of the paper seeks to explain the
strange appearance of image B and the difficulties found in fitting
a lens model to the observed VLBI substructure.
 
\section{Optical and infrared images}

\begin{figure*}
\begin{center}
\includegraphics{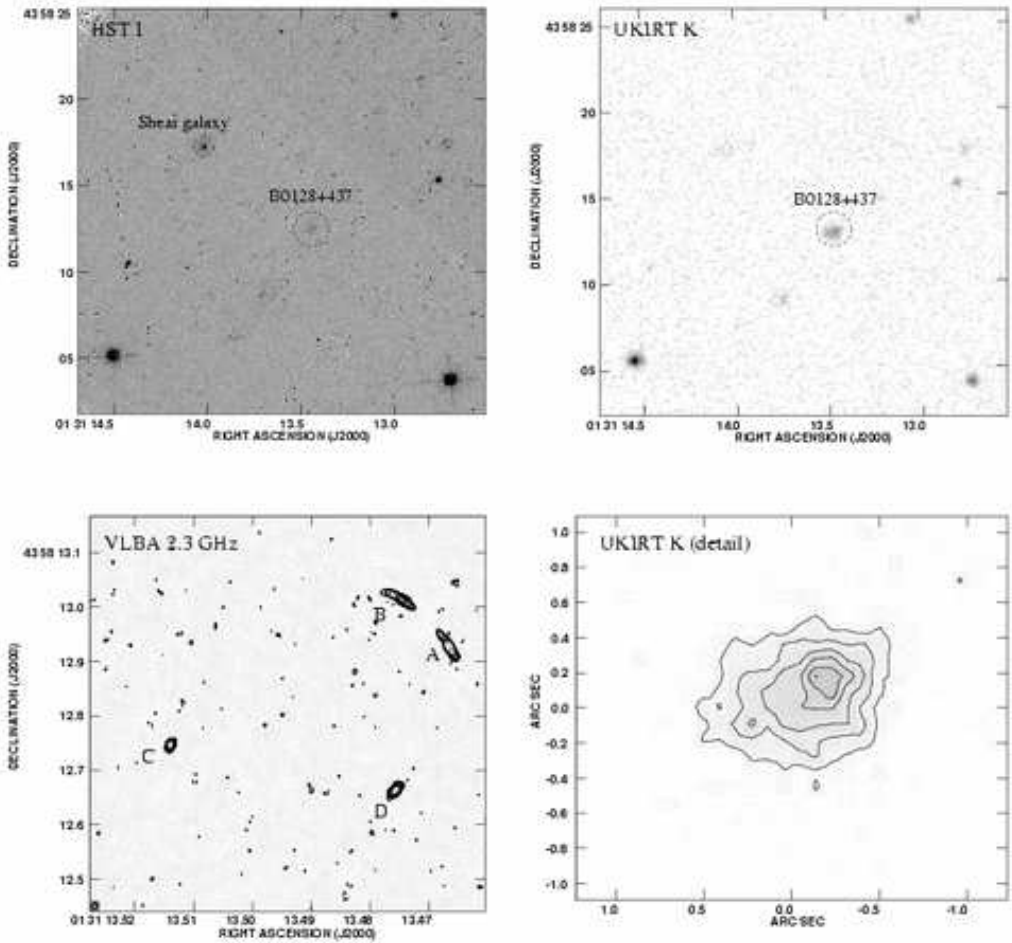}
\caption{Top left: $HST$ $I$-band image with both B0128+437 and the
  possible shear galaxy circled. The axes are labelled according to the
  original $HST$ astrometric information. Top right: UKIRT $K$-band
  (2.2$\mu$m) image with the astrometry calibrated as described in the
  text. The field of view in both images has been restricted to
  $23.5\times23.5$~arcsec$^2$ and each has been aligned using the 
  positions of the two bright stars in the lower half of the
  frame. Bottom left: VLBA 2.3-GHz map (tapered) of B0128+437 that
  illustrates the image configuration. Bottom right: contoured close-up
  of the UKIRT B0128+437 image illustrating the similarity between the
  radio and infrared structure.}
\label{optical}
\end{center}
\end{figure*}

B0128+437 was observed with $HST$ on 2001 June 16 using the Wide Field
and Planetary Camera 2 (WFPC2) giving a spatial sampling of
46~mas~pixel$^{-1}$ at the location of the lens. Data were taken in 
both the F555W and F814W filters which correspond approximately to the
standard Johnson-Cousins $V$ and $I$-bands respectively. Four exposures
were taken in each filter (a two-point dither pattern was used with two
exposures taken at each pointing to enable cosmic ray removal)
resulting in a total exposure time of 5400~s. The final images,
produced using the Image Reduction and Analysis Facility ({\sc iraf}),
show only a very faint and extended detection in $I$-band
($I=23.8\pm0.2$~mag) and none at all in $V$-band ($V>25$). We show the
$I$-band image in Fig.~\ref{optical}. There is no 
evidence of any light from the lensed images and so what light is
detected probably originates in the lensing galaxy. However, the
observations do identify a candidate shear galaxy 7.8~arcsec from
B0128+437 at a position angle of 53$^{\circ}$ (consistent with the
position angle found from the lens modelling of the MERLIN data) with a
total flux of $24.2\pm0.4$~mag in $V$ and $21.8\pm0.1$~mag in $I$.

An image with much higher signal to noise has been obtained with UKIRT
in $K$-band (2.2$\mu$m). These observations were taken on 2000
September 5 using the UKIRT Fast Track Imager (UFTI) which has a plate
scale of 91~mas~pixel$^{-1}$ and a field of view of 93~arcsec. A
standard set of nine 60-s exposures were taken with the telescope
offset by 10~arcsec between exposures. Data reduction was performed
using the ORAC-DR pipeline which corrects the data for bias frames,
bad-pixel masks, dark current subtraction, flat-field correction and
assembles the separate exposures into a final image. This is shown in 
Fig.~\ref{optical} where B0128+437 can be seen as the relatively
bright ($K=17.8\pm0.3$~mag) source in the centre. The source is resolved,
despite the fact that the seeing ($\sim$0.5~arcsec) was approximately
the same as the maximum image separation, with a peak to the north-west
of the centre of the infrared detection. This is best seen in a
contoured close-up view of B0128+437 that is also shown in
Fig.~\ref{optical}. The shear galaxy is extremely faint in this image
and can barely be seen above the noise. 

It is tempting to associate the peak in the infrared image with the two
merging images A and B as the majority of the radio emission is also
located to the north-west of the centre of the system. This would
suggest that the majority of the infrared emission is from the
lensed source, but it is difficult to be sure of this due to the low
resolution and the pointing accuracy of UKIRT which is not sufficient
to allow a direct comparison of the radio and infrared positions. We
have improved on the original UKIRT astrometry by using the measured
positions of 4 stars in the UKIRT field. Three of these stars have
had positions measured using the Carlsberg Meridian Telescope
(D.~W.~Evans, private communication) with the remaining star's position
taken from the Two Micron All Sky Survey (2MASS). The addition of the
2MASS star is useful as it allows us to accurately calibrate the plate
scale along both axes, the Carlsberg stars all lying in a narrow range
of right ascension. The Carlsberg and 2MASS positions of a bright star
in the field differ by only 10~mas showing that the two surveys are
coincident. The astrometry calibration was performed using the STARLINK
{\sc gaia} program and places the UKIRT peak 100~mas to the west of
images A and B, with a much smaller offset in declination. The total
offset is approximately equal to the rms of the astrometric fit
(80~mas) and thus we conclude that the UKIRT peak does correspond to
the radio images A and B. The dominance of the lensed source in the
infrared could stem from the presence of the redshifted
\mbox{[{\sc O\,iii}]} doublet from the source in the passband.

\section{VLBI Observations}

Observations of B0128+437 were made at two separate epochs, the first at
5-GHz using the VLBA only (2000 December 20) and the second simultaneously
at 2.3 and 8.4~GHz using the VLBA and the 100-m Effelsberg antenna
(2002 January 2). On both occasions the source was 
observed for a total of $\sim$14~h, utilising fast phase-referencing between
the weak target and the nearby (37~arcmin distant) calibrator
J0132+4325 with a cycling time of 4.5~min. Both epochs employed a data
recording rate of 128~Mb~s$^{-1}$ with 2-bit sampling. This allowed, at
5~GHz, two 
contiguous 8-MHz bands (IFs) to be recorded, both with left (LCP) and 
right (RCP) circular polarisations. As the 2.3 and 8.4~GHz data were
observed simultaneously, two contiguous 8-MHz bands were assigned to
each frequency and only RCP was recorded. The data were subsequently
processed using the VLBA correlator in Socorro, producing 16 0.5-MHz
channels per IF and data averaged into 2-s bins. Cross-polarised
visibilities were not produced as the MERLIN 5-GHz data show the images
to be unpolarised.

The data were calibrated and mapped using the NRAO {\sc aips} software
package. Once the flux scale was set using measurements of system
temperature and {\em a priori} gain curves, phase slopes across the
band were removed using a short segment of data on a bright source. The
phase-reference source was then fringe-fitted over the entire length of
the experiment, solving for phase, phase rate and delay. The solutions
so found were then transferred to the target. Secondary amplitude
solutions were also derived from the phase calibrator. An additional
step was undertaken at 2.3~GHz where the rapid phase variations
caused by the ionosphere significantly degraded the success of the
phase referencing. We therefore used the {\sc aips} task {\sc
  tecor} which attempts to remove the effects of the ionosphere using
maps of the vertical Total Electron Content (vTEC) of the atmosphere;
these are produced by a network of Global Positioning System (GPS)
receivers. Application of the {\sc tecor} corrections significantly
slowed the time variation of the phases of the phase-reference source
and produced maps of B0128+437 with a higher dynamic range and with
structures that were much more plausible given the prior knowledge we
had from the 5-GHz data taken a year previously.

The data were then averaged in frequency from 16 channels per band to
4 and in time from 2~sec into 10-sec samples. In order to prevent the
VLBA-Effelsberg baselines from dominating the final image, the
visibility data weights were altered by taking their fourth
root. Iterative mapping and self-calibration was then undertaken using
the AIPS tasks {\sc imagr} and {\sc calib}, four maps being made
at the location of each image as measured from the MERLIN 5-GHz
map. At each frequency rms off-source noise levels were achieved that
were close to the theoretical values.

\begin{figure*}
\begin{center}
\includegraphics{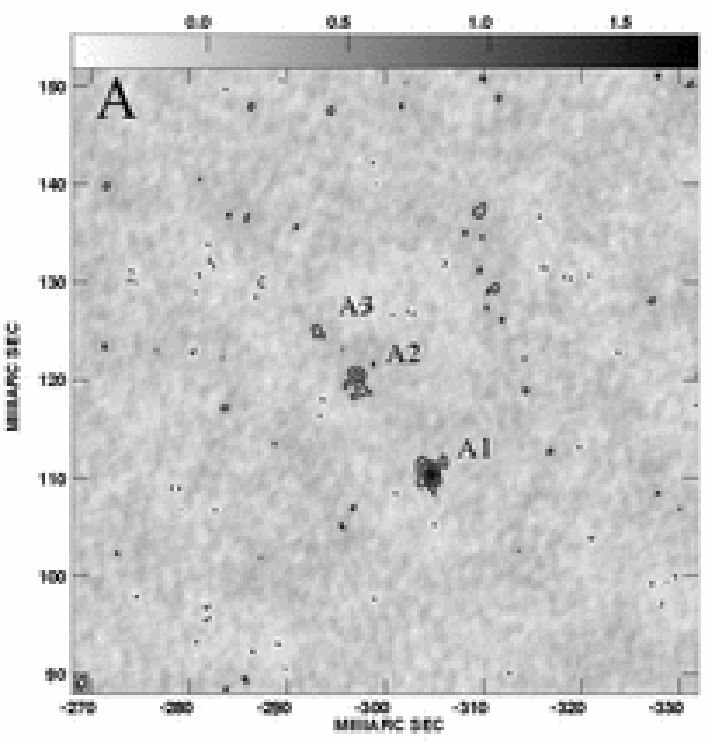}
\includegraphics{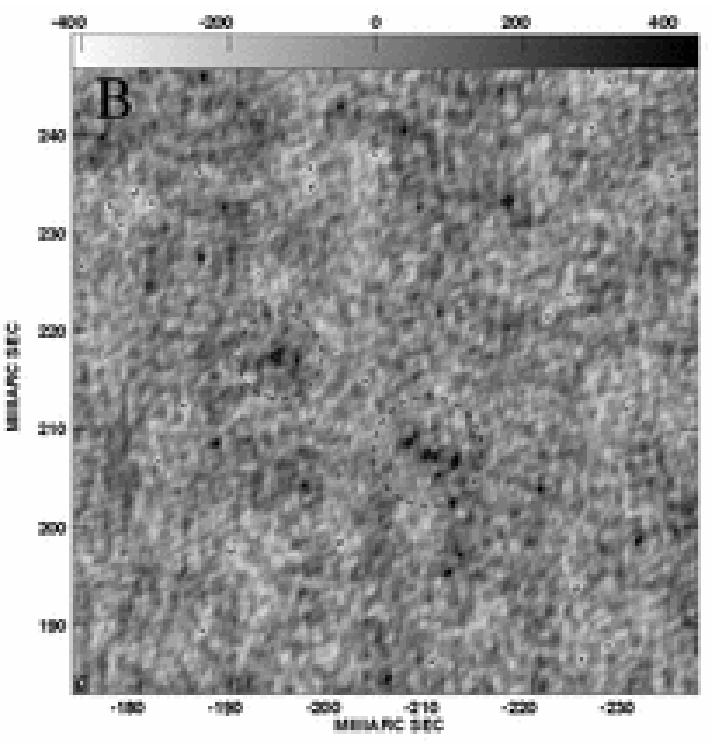}
\includegraphics{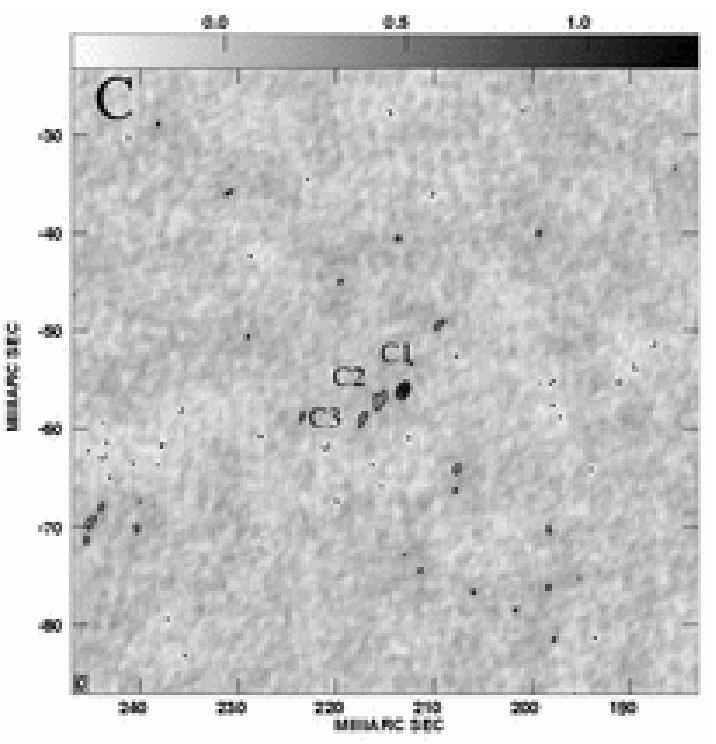}
\includegraphics{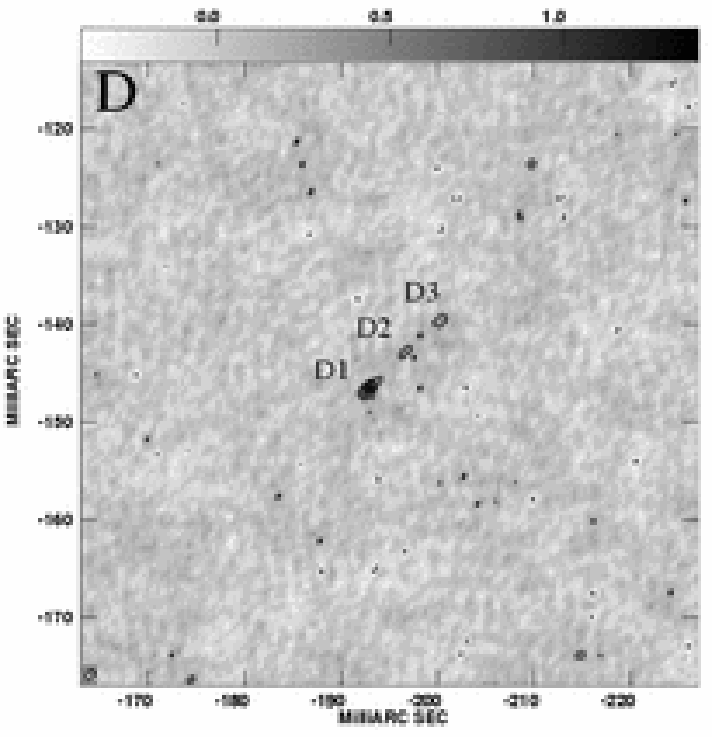}
\caption{Naturally-weighted VLBA+Effelsberg maps (RCP) of CLASS B0128+437
  at 8.4~GHz. Contours are plotted at
  multiples ($-1$, 1, 2, 4, 8, etc) of 3$\sigma$ where $\sigma$ is the
  off-source rms noise in the map (90~$\mu$Jy~beam$^{-1}$). The restoring
  beam is shown in the bottom-left corner and has a FWHM of $1.2 \times
  0.7$~mas$^2$ at a position angle of $-21$\fdg6. The grey scales represent
  surface brightness in units of mJy\,beam$^{-1}$, apart from the map of
  image B where the units are $\mu$Jy\,beam$^{-1}$. Labels that identify
  the corresponding sub-components in images A, C and D are plotted
  above. In the case of image B we plot two dashed circles marking the
  location of what faint emission there is. The origin used for these
  and subsequent maps is $01^{\rmn{h}} 31^{\rmn{m}} 13\fs494, +43\degr
  58\arcmin 12\farcs805$ (J2000).}
\label{mapsx}
\end{center}
\end{figure*}

\begin{figure*}
\begin{center}
\includegraphics{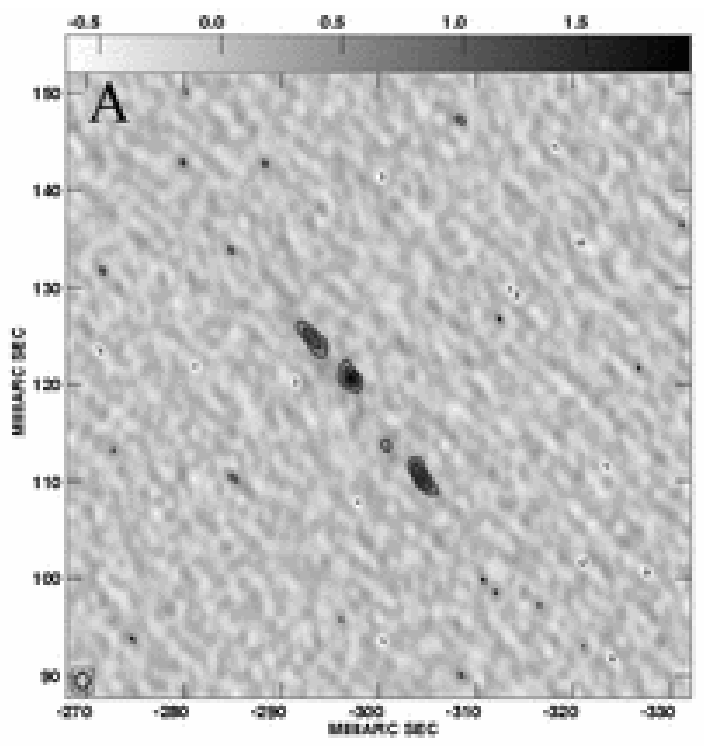}
\includegraphics{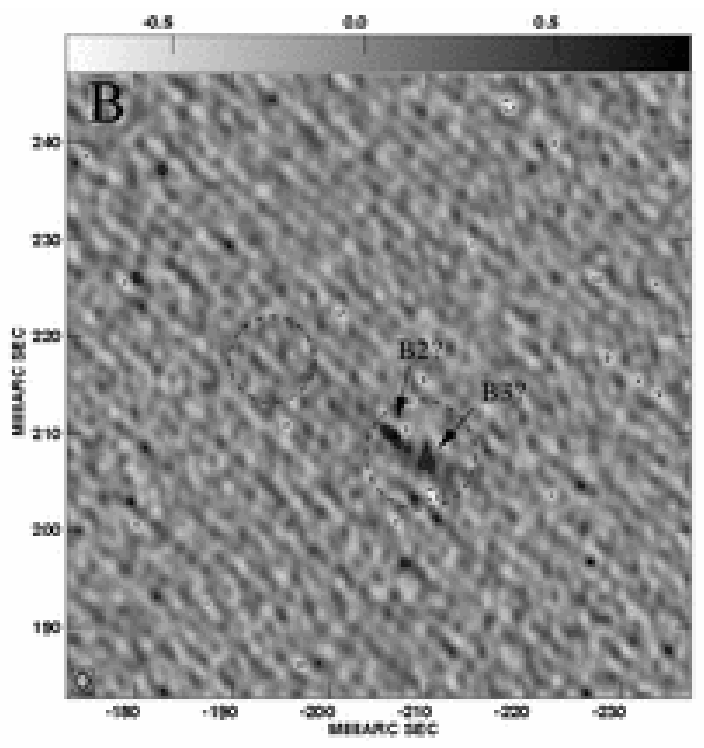}
\includegraphics{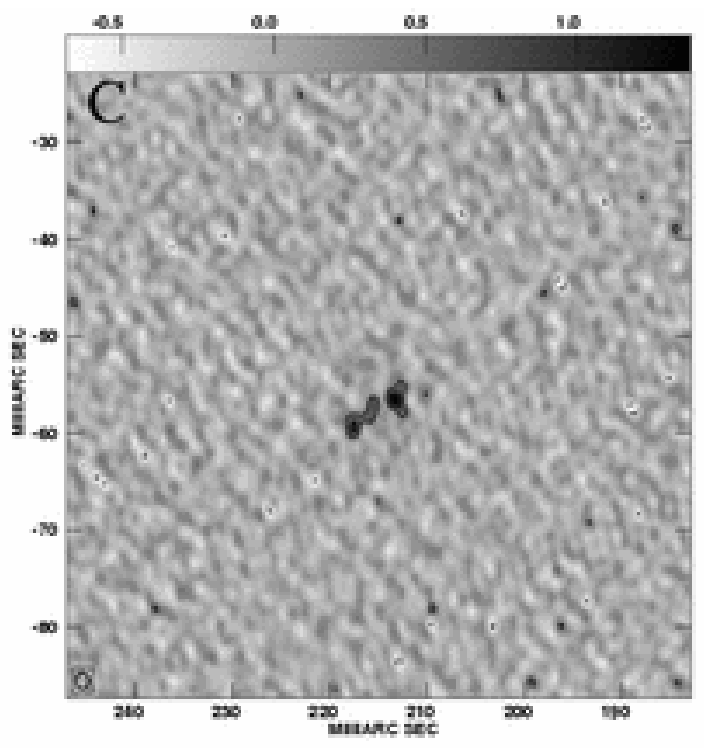}
\includegraphics{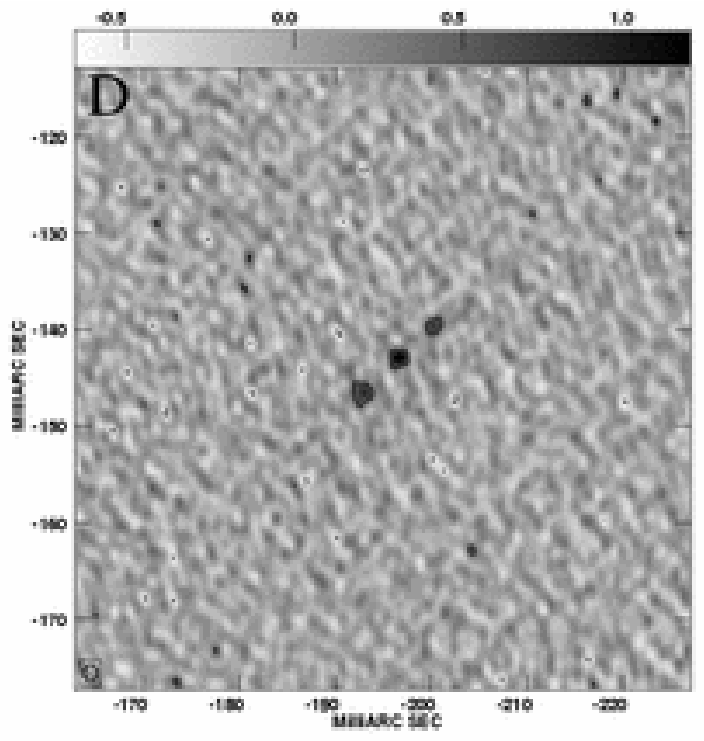}
\caption{Uniformly-weighted VLBA maps (Stokes $I$) of CLASS B0128+437 at
  5~GHz. Contours are plotted at
  multiples ($-1$, 1, 2, 4, 8, etc) of 3$\sigma$ where $\sigma$ is the
  off-source rms noise in the map (155~$\mu$Jy~beam$^{-1}$). The restoring
  beam is shown in the bottom-left corner and has a FWHM of $1.7 \times
  1.3$~mas$^2$ at a position angle of $0$\fdg6. The grey scales represent
  surface brightness in units of mJy\,beam$^{-1}$. The maps are plotted
  on the same angular scale as the 8.4-GHz maps in Fig.~\ref{mapsx}. The
  same dashed circles that appear on the 8.4-GHz map of image B are
  reproduced here as a guide to the eye. We also label the possible
  identifications of sub-components B2 and B3 (see
  Section~\ref{discussscattering}).}
\label{mapscuni}
\end{center}
\end{figure*}

\begin{figure*}
\begin{center}
\includegraphics{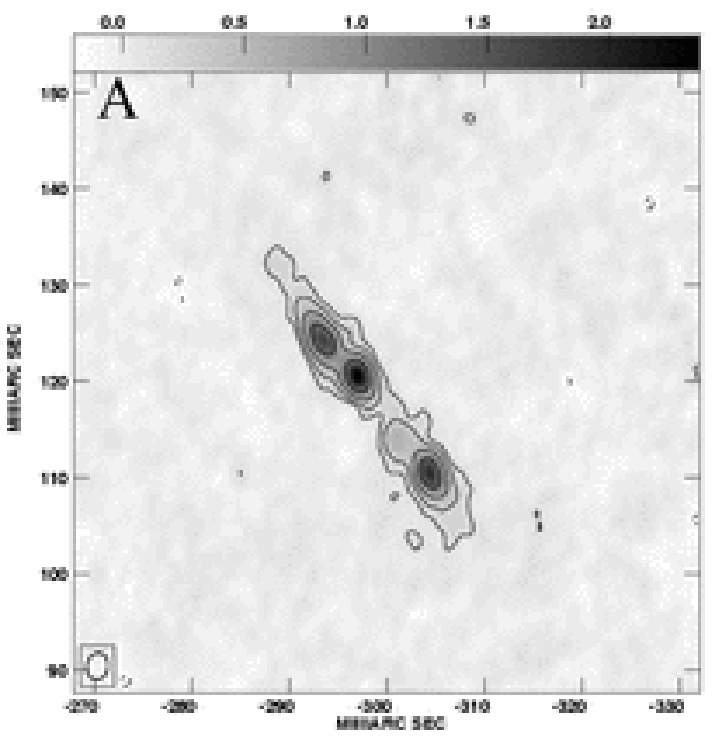}
\includegraphics{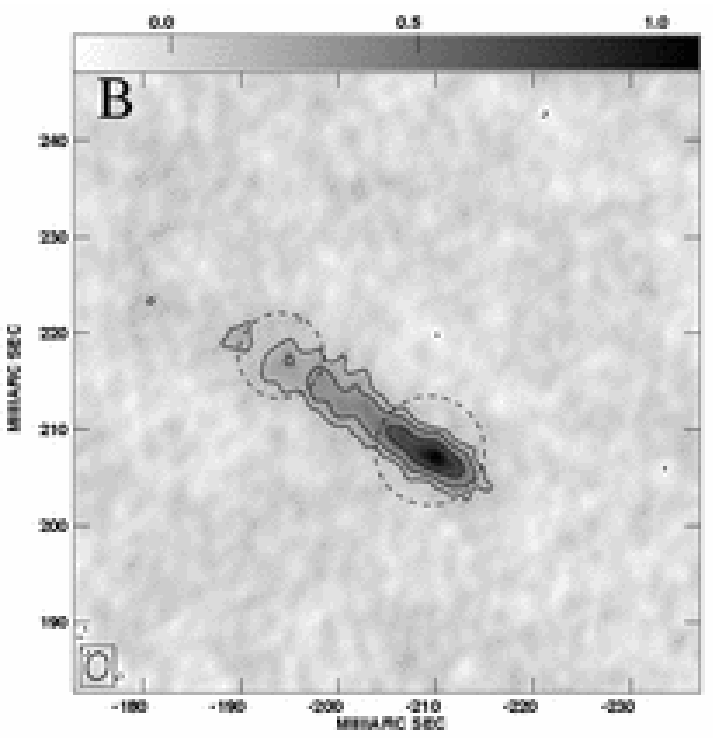}
\includegraphics{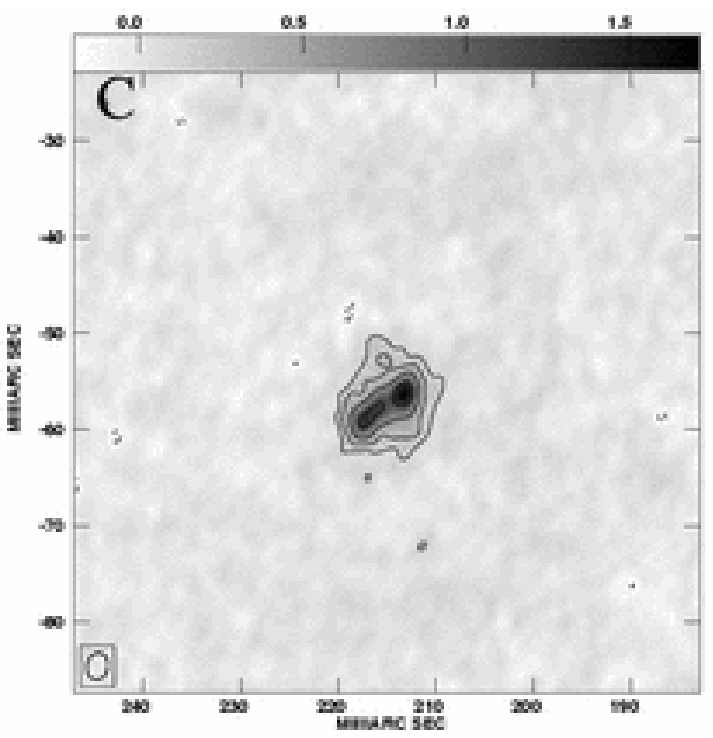}
\includegraphics{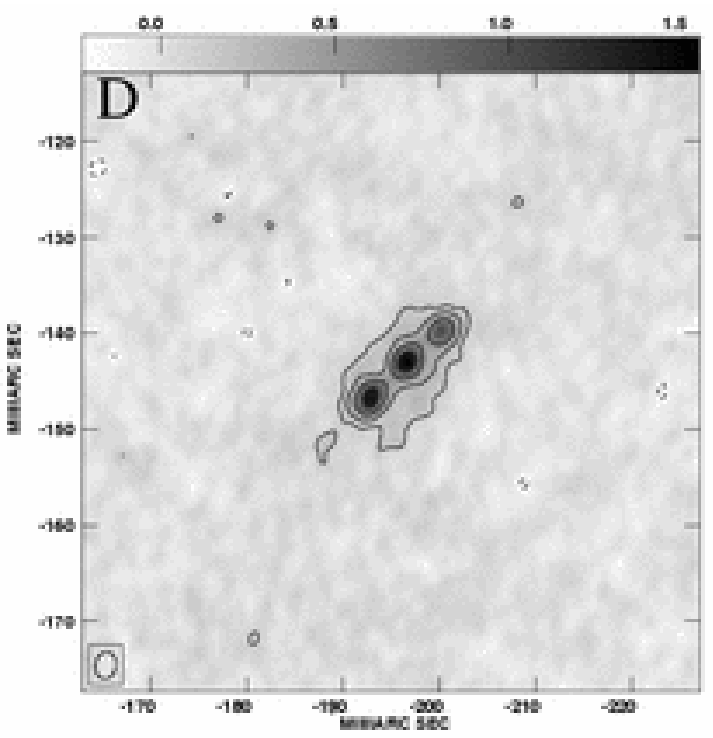}
\caption{Naturally-weighted VLBA maps (Stokes $I$) of CLASS B0128+437 at
  5~GHz. Contours are plotted at
  multiples ($-1$, 1, 2, 4, 8, etc) of 3$\sigma$ where $\sigma$ is the
  off-source rms noise in the map (50~$\mu$Jy~beam$^{-1}$). The restoring
  beam is shown in the bottom-left corner and has a FWHM of $2.8 \times
  2.2$~mas$^2$ at a position angle of $-10$\fdg4. The grey scales represent
  surface brightness in units of mJy\,beam$^{-1}$. The maps are plotted
  on the same angular scale as the 8.4-GHz maps in
  Fig.~\ref{mapsx}. The same dashed circles that appear on the 8.4-GHz
  map of image B are reproduced here as a guide to the eye.}
\label{mapsc}
\end{center}
\end{figure*}

\begin{figure*}
\begin{center}
\includegraphics{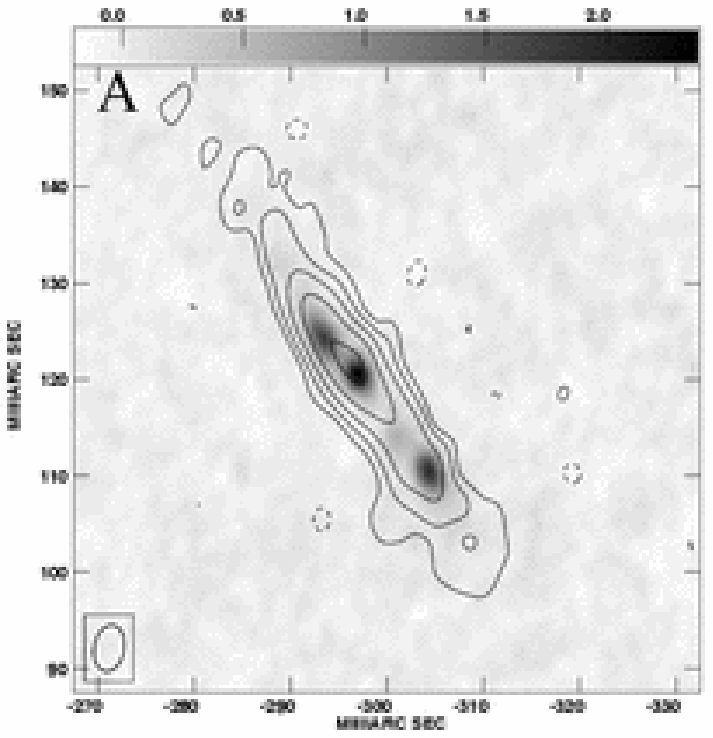}
\includegraphics{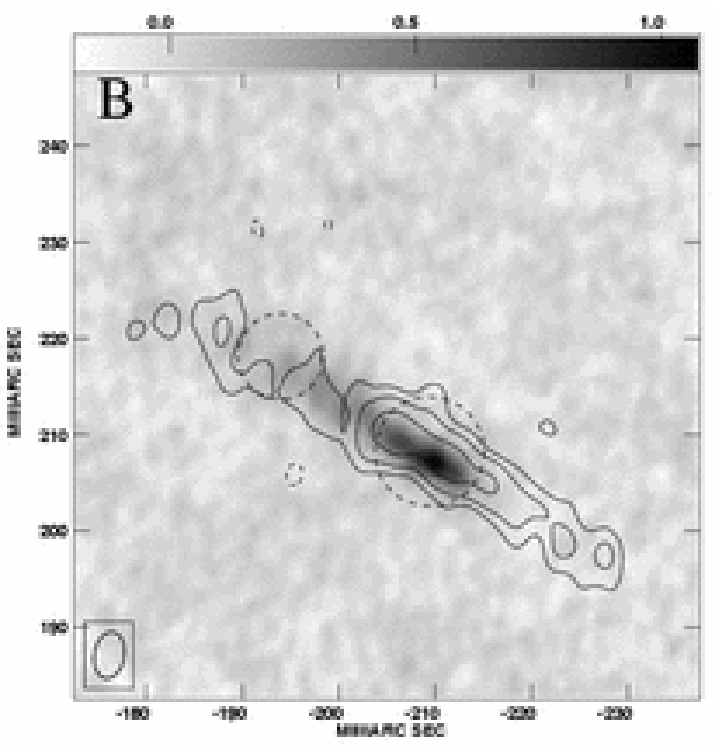}
\includegraphics{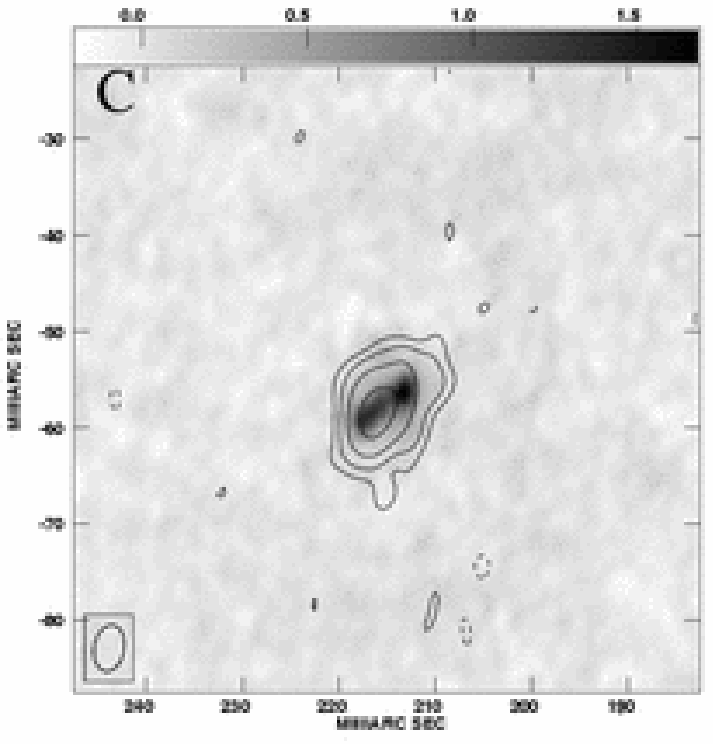}
\includegraphics{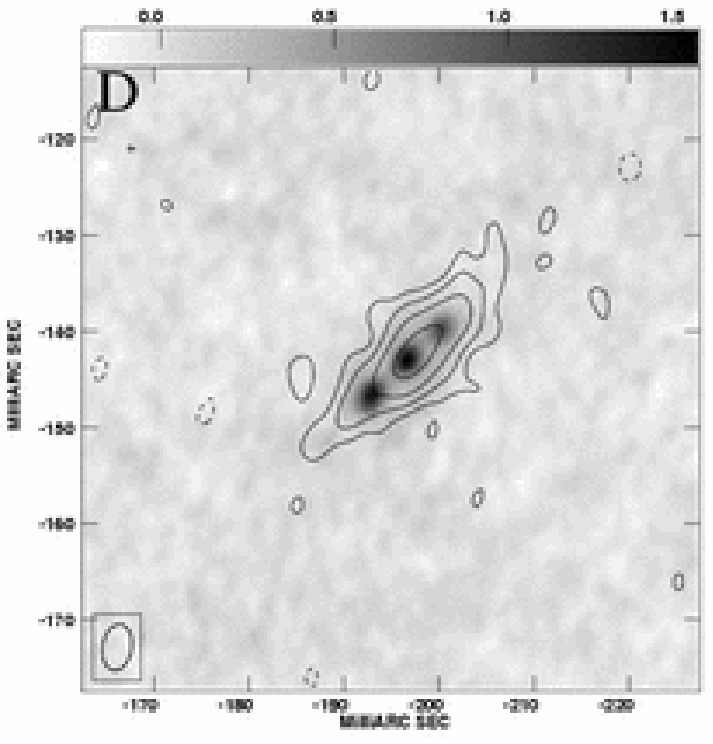}
\caption{VLBA maps (RCP) of CLASS B0128+437 at 2.3~GHz (contours)
  made with a robust 0 weighting scheme. Contours are plotted at
  multiples ($-1$, 1, 2, 4, 8, etc) of 3$\sigma$ where $\sigma$ is the
  off-source rms noise in the map (100~$\mu$Jy~beam$^{-1}$). The restoring
  beam is shown in the bottom-left corner and has a FWHM of $4.8 \times
  3.1$~mas$^2$ at a position angle of $-10$\fdg1. The grey scales show the
  5-GHz surface brightness distribution in units of mJy\,beam$^{-1}$.
  The maps are plotted on the same angular scale as the 8.4-GHz maps in
  Fig.~\ref{mapsx} and the 5-GHz maps in Fig.~\ref{mapsc}. The same
  dashed circles that appear on the 8.4-GHz map of image B are again
  reproduced here.}
\label{mapss}
\end{center}
\end{figure*}

\section{VLBI maps}
\label{vlbimapsec}

The final maps are shown in Figs~\ref{mapsx}, \ref{mapscuni},
\ref{mapsc} and \ref{mapss} and reveal a wealth of image
substructure. If we consider first only the 8.4-GHz maps, we see that
images A, C and D all consist of three discrete and approximately
colinear sub-components. We have labelled these A1, A2, A3, etc in
Fig.~\ref{mapsx}. Correctly labelling each sub-component within each
image is made particularly easy due to the fact that one of the
sub-components (1) is much brighter than the 
others. This ensures that there is no ambiguity over which 
sub-component in each image corresponds to the same part of the
source. In contrast, image B looks significantly different to the other
images and is in fact barely detected. There are though two distinct
areas of emission (which we have highlighted with dashed circles in
Fig.~\ref{mapsx}) which we identify with the lensed image despite their
very low surface brightness ($\sim3\sigma$). The circles in
Fig.~\ref{mapsx} are also reproduced in the 5 and 2.3-GHz maps.

At 5~GHz a very similar picture holds with images A, C and D still
dominated by three compact components. This is seen most clearly in the
maps of Fig.~\ref{mapscuni} where the use of uniform weighting (where
the nominal visibility weights are modified by the local density of data
points in Fourier space) emphasises the longer baselines, thus producing
the highest possible resolution. Image B is detected more securely here
although the image remains very weak. The emission that is detected
lies within the western of the two circles that indicate the location
of probable lensed emission at 8.4~GHz and appears to be concentrated
in two small patches, separated by about 5~mas. There is no evidence of
any emission within the easternmost circle. However, it is not
only image B that is anomalous at this frequency as the relative
brightnesses of the sub-components in image C do not agree with those
seen in images A and D. In these last two images the central
sub-component (2) has a higher surface brightness than the other two,
contrasting with the situation at 8.4~GHz where sub-component 1 is the
brightest. In image C though the
central sub-component has the lowest surface brightness. A final
interesting feature of these maps is the obvious stretching of
sub-components A1 and A2 along the direction of the jet that is not
seen in image D.

We have also made maps of the images at 5~GHz using natural weighting
i.e. without a correction for the density of visibility samples. These
maps, shown in Fig.~\ref{mapsc}, are sensitive to larger-scale
structures than those detected with uniform weighting and faint extended
emission is detected in all images e.g. along the jet axis in image A
whilst C and D appear to sit in faint haloes. However, whilst these images
appear similar to their uniformly-weighted counterparts,
image B looks dramatically different. Instead of the few weak patches of
emission in the uniformly-weighted map, this image is now detected with
high signal to noise and is recognisable as an approximately straight
jet with a length of $\sim$30~mas. The peak surface brightness lies at
the south-western end of the jet which fades gradually towards the
north-east.

We have fitted elliptical Gaussians to the 5-GHz visibility data using
the {\sc omfit} task in {\sc aips}. We fitted nine Gaussians in total,
one for each sub-component in images A, C and D; image B is removed
from the data by subtracting its CLEAN components found during the
naturally-weighted deconvolution of the data. In calculating the 
errors it is assumed that the model is a perfect representation of the
data, {\sc omfit} scaling the weights attached to the visibility data
(which are unknown in an absolute sense) until the $\chi^2$ per degree
of freedom ($\overline{\chi}^2$) of the modelfit is equal to unity. As
the images are dominated by compact emission we believe that this is a
fairly safe assumption. The results of the modelfit are shown in
Table~\ref{modelfit}. Repeated trials, with different starting models,
established that there is much more leeway in fitting parameters to
image C than to the other images, presumably due to it being less well
resolved. We therefore consider the model fits for image C to be less
reliable than for the other images. Table~\ref{modelfit} also shows the
results of model fitting to sub-components 1 only of the 8.4-GHz data,
again using {\sc omfit}. This demonstrates that sub-components 2 and 3
have steeper spectra than 1 as this has a similar flux density at both
frequencies whilst Figs~\ref{mapsx}, \ref{mapscuni} and \ref{mapsc}
show that the other two significantly brighten. A comparison between
the positions measured at 5 and 8.4~GHz indicate that the phase
referencing has been very successful, giving positions for
sub-component 1 that are the same to within 0.2~mas.

At the lowest frequency, 2.3~GHz, all the images look smoother and
larger (Fig.~\ref{mapss}) with little sign of the compact structures
that dominate the images at the higher frequencies. This is due to a
combination of the lower resolution and the increasing dominance of
both the steep-spectrum sub-components (2 and 3) and the surrounding
extended jet emission at the lower frequency. There is, though, evidence
that the sub-components are being significantly resolved at 2.3~GHz as
there is a lack of correlated flux on the longest baselines. This
became apparent during phase self-calibration when practically no
solutions could be found for Effelsberg, the most sensitive antenna in
the array and the one contributing the longest baselines. This
telescope was therefore flagged and does not contribute to the maps in
Fig.~\ref{mapss}; a map made using only the baselines to Effelsberg
(sensitive to emission on scales of 3--4~mas) contains nothing but
noise.

Finally, image B looks much more like image A at 2.3~GHz and it now
becomes apparent that these images do point towards each other, as
expected from lens models (see Section~\ref{modelsec}). However,
inconsistencies between image B and image A remain, the most obvious
being that the surface brightness of image B drops off very sharply
about halfway along its length (at around $-$200, 210~mas, between the
two dashed circles) with the remaining half of the image barely
detected. The correspondence in Fig.~\ref{mapss} between the contours
and greyscales, which represent the 5-GHz images, demonstrate that the 
phase-referencing has also been successful at 2.3~GHz, despite the
problems introduced by the ionosphere.

\begin{table*}
\begin{center}
\caption{Flux densities, positions and shapes of the compact
  sub-components identified in the VLBI observations, derived
  from the {\sc{aips}} task {\sc{omfit}}. The top half of the table
  gives the results for the three sub-components of A, C and D derived
  from the 5-GHz data whilst the bottom half gives results for the
  8.4-GHz data, for sub-component 1 only. {\sc omfit} was unable to
  determine values for parameters marked with a $-$. Position
  angles are measured North through East and positions are again offset
  from $01^{\rmn{h}} 31^{\rmn{m}} 13\fs494, +43\degr 58\arcmin
  12\farcs805$ (J2000).}
%\newcolumntype{d}{D{\pm}{\pm}{-1}}
\begin{tabular}{cr@{$\pm$}lr@{$\pm$}lr@{$\pm$}lr@{$\pm$}lr@{$\pm$}lr@{$\pm$}l} \hline
Sub-component & \multicolumn{2}{c}{Flux density (mJy)} & \multicolumn{2}{c}{East offset (mas)} & \multicolumn{2}{c}{North offset}
(mas) & \multicolumn{2}{c}{Major axis (mas)} & \multicolumn{2}{c}{Axial ratio} & \multicolumn{2}{c}{Position angle ($^{\circ}$)} \\ \hline 
A1 & 3.9&0.1 & $-$304.3&0.1 &  110.4&0.1  & 4.2&0.2 & 0.27&0.04 &
27.0 & 2.1   \\
A2 & 3.1&0.2 & $-$297.1&0.0 &  120.6&0.1  & 2.5&0.2 & 0.20&0.09 & 
28.0 & 2.9   \\
A3 & 4.0&0.2 & $-$293.4&0.1 &  124.4&0.1  & 5.8&0.4 & 0.16&0.03 &
30.8 & 1.5   \\
C1 & 2.7&0.1 & 213.2&0.1  &  $-$56.3&0.1  & 3.1&0.2 & 0.42&0.07 &
$-$10.1 & 4.4  \\
C2 & 1.2&0.1 & 215.6&0.1  &  $-$57.5&0.1  & 1.9&0.4 & \multicolumn{2}{c}{$-$} &
$-$9.2 & 7.7  \\
C3 & 1.6&0.1 & 217.5&0.1  &  $-$59.3&0.1  & 2.0&0.3 & 0.33&0.14 &
$-$14.1 & 7.2 \\
D1 & 2.4&0.1 & $-$193.0&0.1 &  $-$146.7&0.1 & 3.5&0.3 & 0.34&0.06 &
$-$46.6 & 3.9  \\
D2 & 2.0&0.1 & $-$196.6&0.1 &  $-$142.9&0.1 & 1.7&0.2 & 0.40&0.14 &
$-$58.3 & 8.5  \\
D3 & 1.4&0.1 & $-$200.1&0.1 &  $-$139.7&0.1 & 2.0&0.3 & 0.42&0.18 &
$-$69.8 & 11.4  \\ \hline
A1 & 3.6&0.3 & $-$304.5&0.1 &  110.4&0.1  & 1.6&0.2 & 0.31&0.11 &
29.0 & 6.5   \\
C1 & 1.3&0.2 & 213.1&0.0  &  $-$56.1&0.1  & 0.4&0.3 & \multicolumn{2}{c}{$-$} &
$-$35.1 & 37.3    \\
D1 & 2.2&0.2 & $-$192.8&0.1 &  $-$146.7&0.1 & 1.1&0.2 & 0.20&0.22 &
$-$55.0 & 8.0    \\ \hline
\end{tabular}
\label{modelfit}
\end{center}
\end{table*}

\section{Lens modelling}
\label{modelsec}

Our aim is to see if a lens model with a smooth mass distribution can
fit the many new observational constraints now available.
\citet{phillips00} were able to construct a model for the mass
distribution in the lens galaxy despite having no optical information
regarding the position and light distribution in the galaxy. They found
that a singular isothermal ellipsoid (SIE) model
\citep*[e.g.][]{kormann94} including external shear was able to recover
the MERLIN image positions and flux density ratios to within 1~mas and
a few per~cent respectively. In this paper we attempt to improve the
modelling by using the same parameterisation, but incorporating the
VLBI image constraints presented in Section~\ref{vlbimapsec}. We have
also experimented with the novel Fourier method of \citet{evans03}.

\subsection{SIE+external shear parameterisation}

In order to be as rigorous as possible, we have performed the SIE
modelling using three separate codes. Two of these are 
written by ourselves, {\sc igloo} (NJJ) and {\sc glint} (TY), as well
as the {\sc lensmodel} package \citep{keeton03}. The results presented
in this section all use full image-plane optimisation for the greatest
accuracy. We find that the results of the modelling are broadly
independent of the software used, although there are small differences.
Chi-squareds, for example, can differ by up to a factor of a few and
slightly different values for the galaxy parameters can be obtained,
especially those that are degenerate. At no point do we use the flux
densities to constrain the model as these may be affected by scattering
in the lens galaxy (see Section~\ref{discussscattering}). The models
are still well constrained, though, due to the large number of degrees
of freedom provided using the positional constraints alone.

Determining the image positions for image B is of course difficult due
to the fact that the three sub-components are not well-defined in this
image. We are forced therefore to use larger errors for this image. As
inputs to the model we sometimes used the same position for all three
sub-components, a point close to the centre of the jet. On other
occasions we used our best estimates of the positions of the B
sub-components from the VLBI maps (see
e.g. Section~\ref{discussscattering}). In both cases we assumed large
(5-10~mas) errors on the positions.

\begin{figure*}
\begin{center}
\includegraphics[scale=0.33,bb=50 50 554 554,angle=-90]{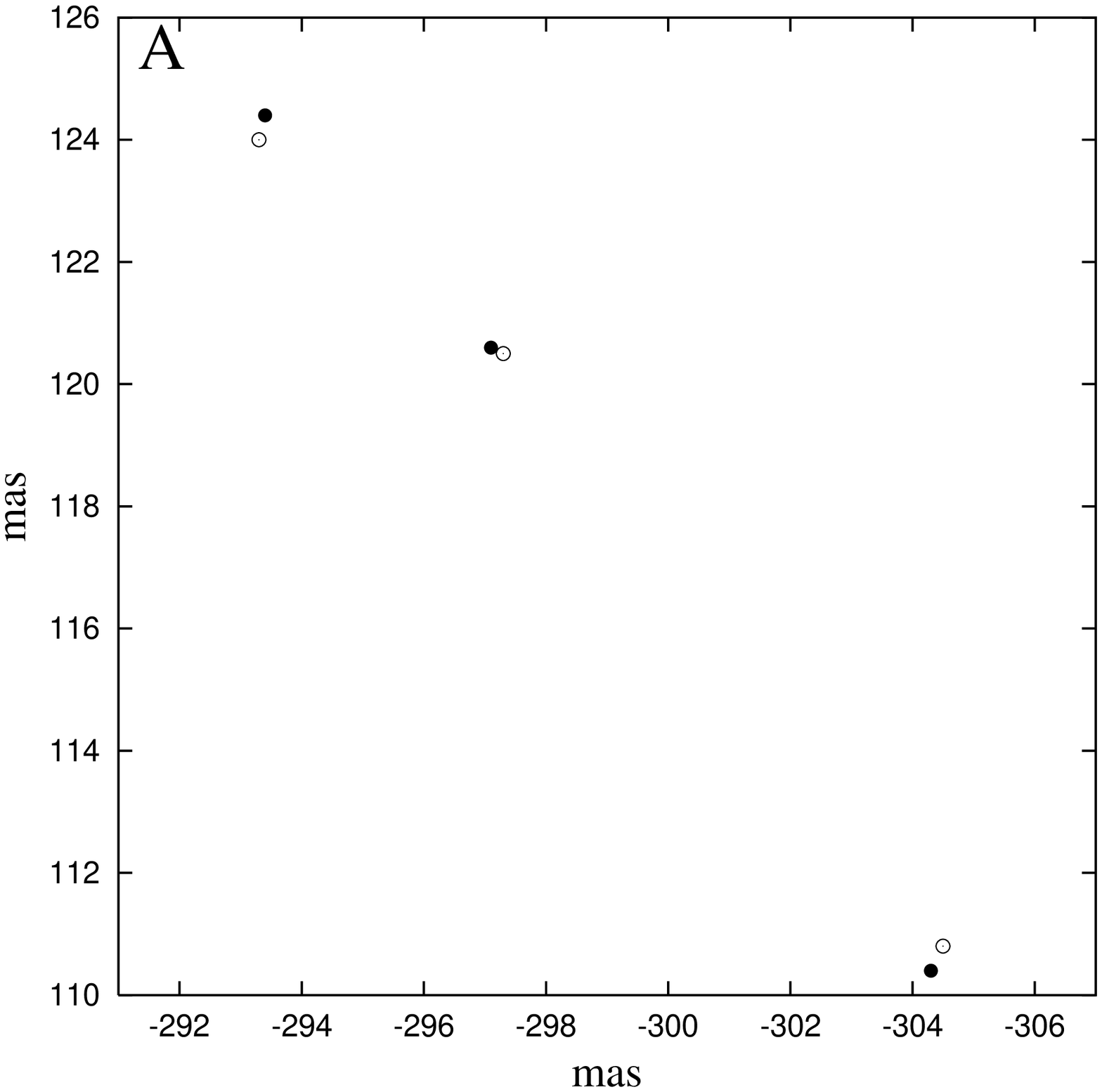}
\includegraphics[scale=0.33,bb=50 50 554 554,angle=-90]{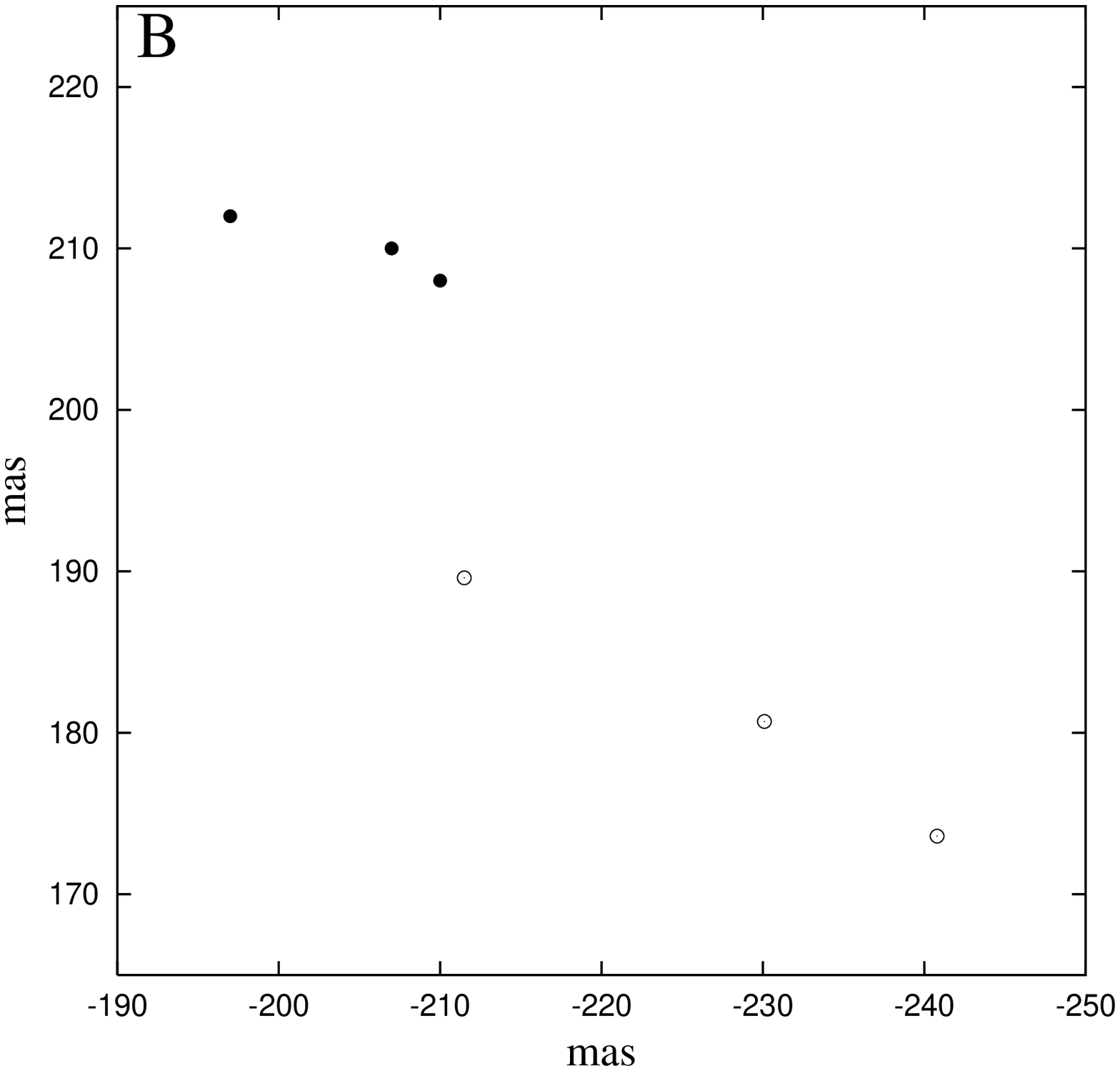}
\includegraphics[scale=0.33,bb=50 50 554 554,angle=-90]{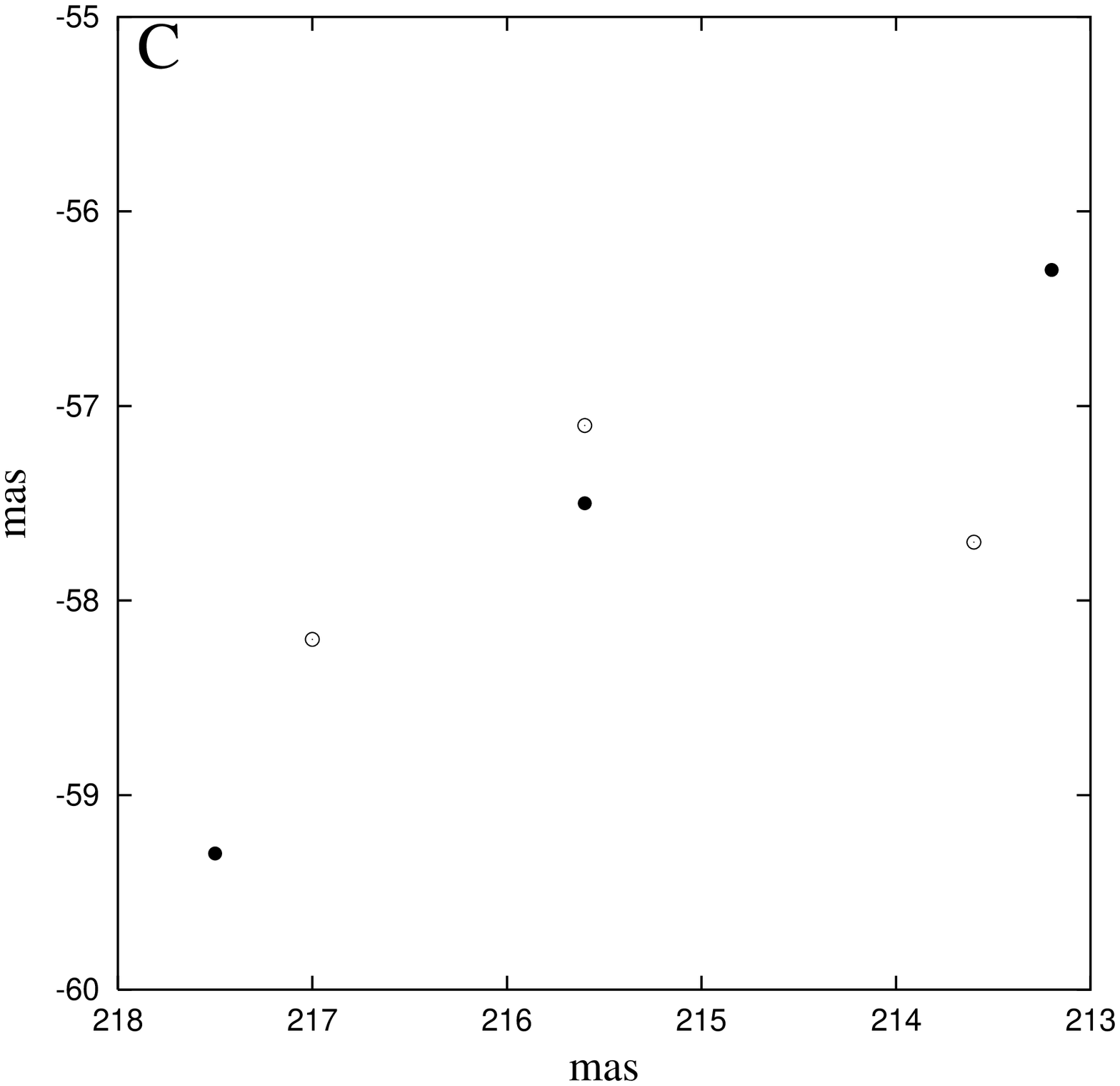}
\includegraphics[scale=0.33,bb=50 50 554 554,angle=-90]{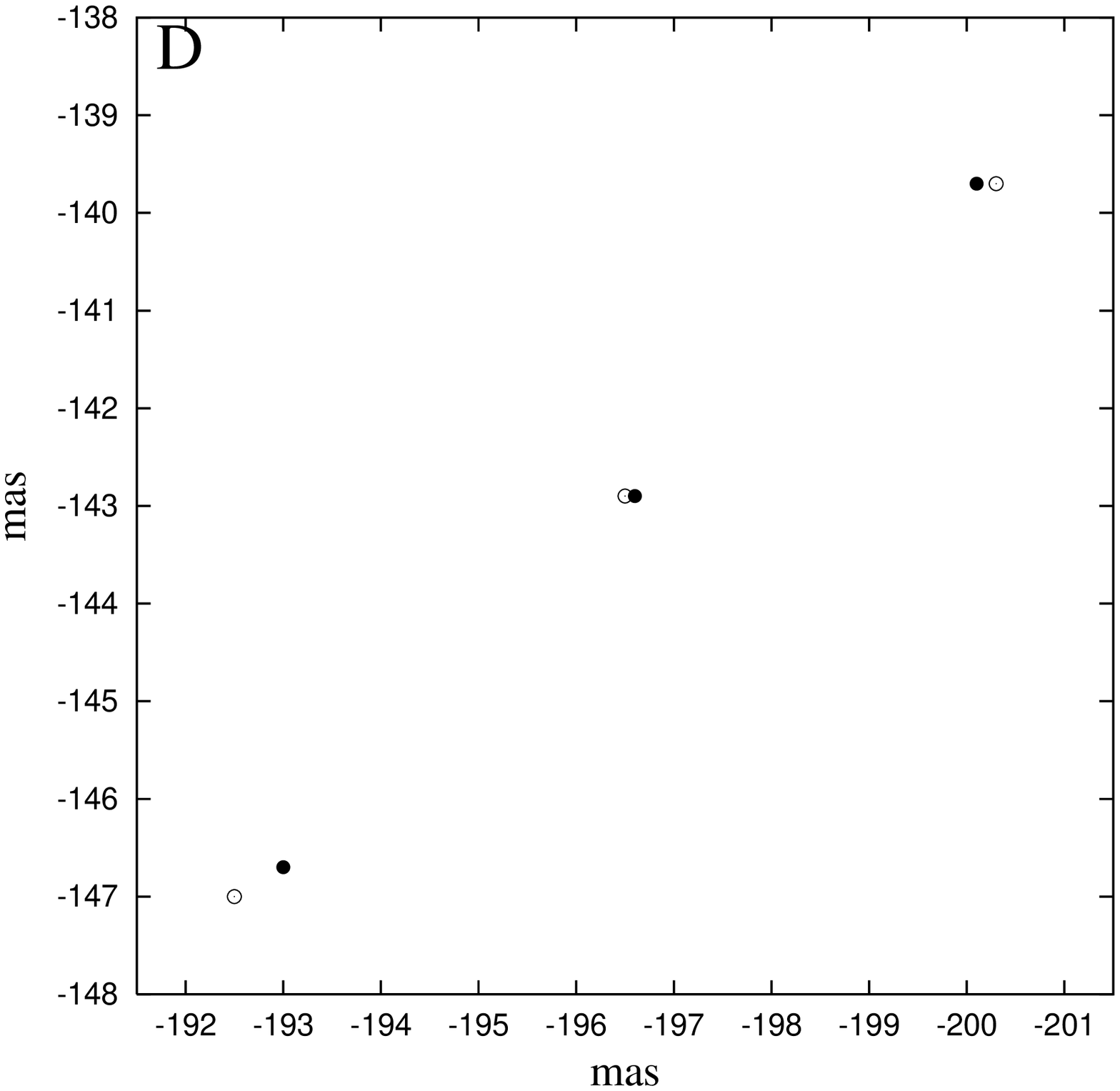}
\caption{Observed positions of image sub-components (filled circles)
  and derived model positions (open circles) for Model 1 using the {\sc
  glint} lens modelling software. The image B positional constraints
  have been estimated from the VLBI maps.}
\label{model1}
\end{center}
\end{figure*}

We begin by assuming that the true errors on the A, C and D
sub-component image positions are those found during the model fitting
to the observational data i.e. 0.1~mas. This we will refer to as
Model~1, with 5~mas errors assumed for the positions of the image B
sub-components. The fit found using Model~1 is extremely poor, the
chi-squared being equal to 635. With 11 degrees of freedom (24
positional constraints and 13 model parameters) this corresponds to a
reduced chi-squared of 58. The reason for the high $\chi^2$ can be seen
in Fig.~\ref{model1} where we plot the observed image positions along
with those found after optimising the model. Image B is fit very poorly
(lying $\sim$30--40~mas away from its observed position) as is image C,
the modelled sub-components C1 and C3 suffering from offsets of
$>$1~mas. The parameters of the optimised model are shown in
Table~\ref{lensmodtab}.

\begin{figure*}
\begin{center}
\includegraphics[scale=0.33,bb=50 50 554 554,angle=-90]{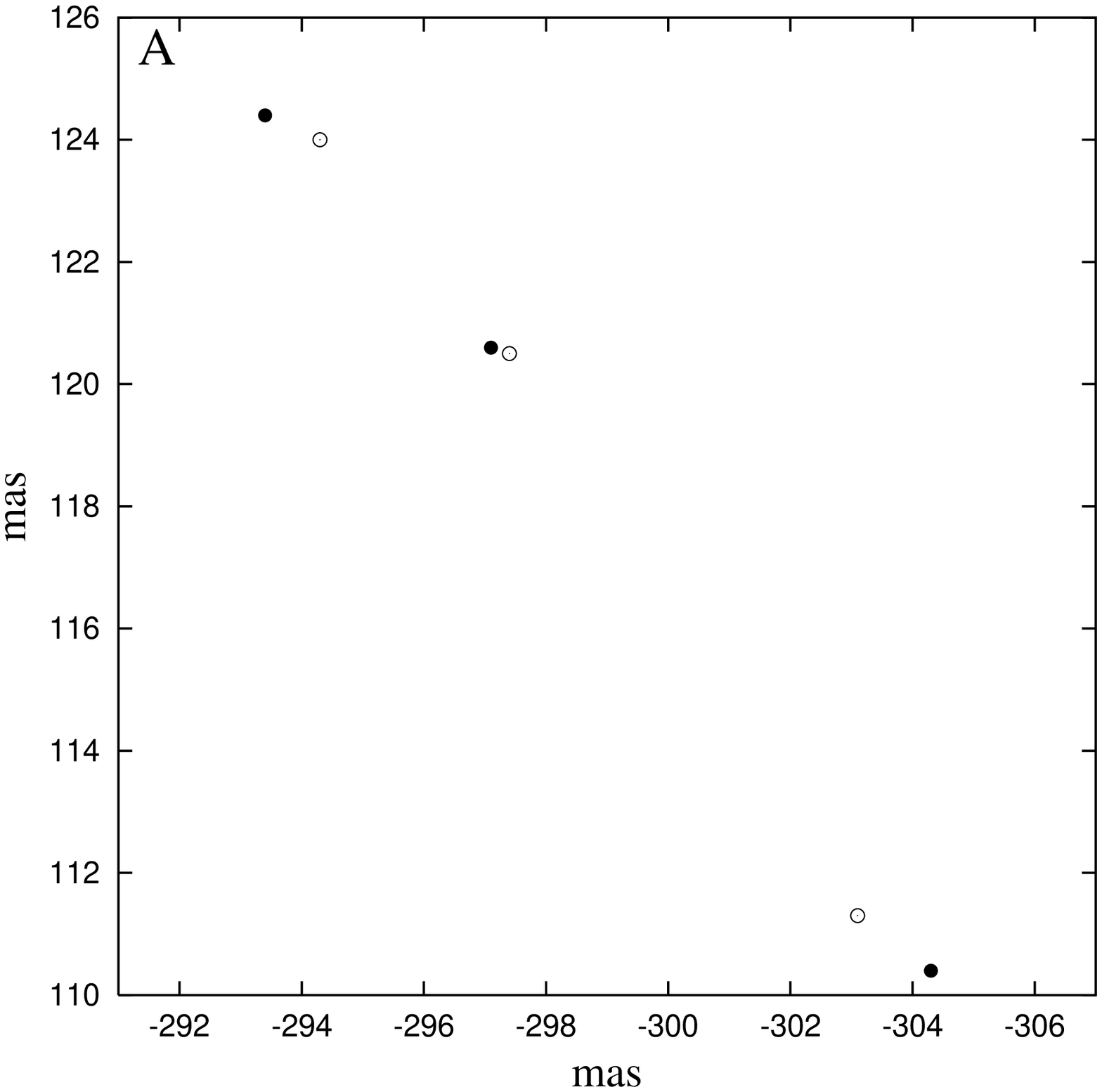}
\includegraphics[scale=0.33,bb=50 50 554 554,angle=-90]{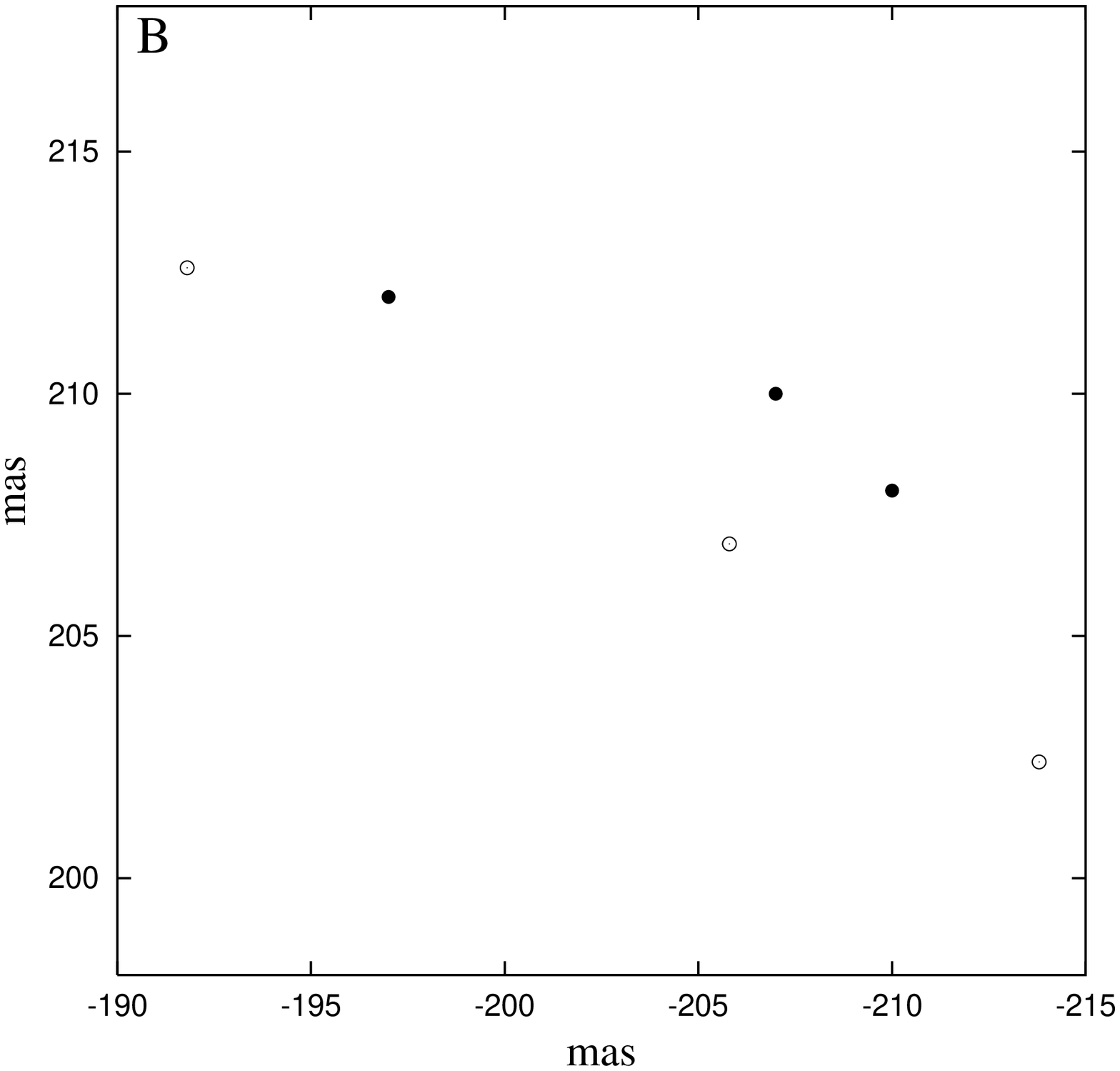}
\includegraphics[scale=0.33,bb=50 50 554 554,angle=-90]{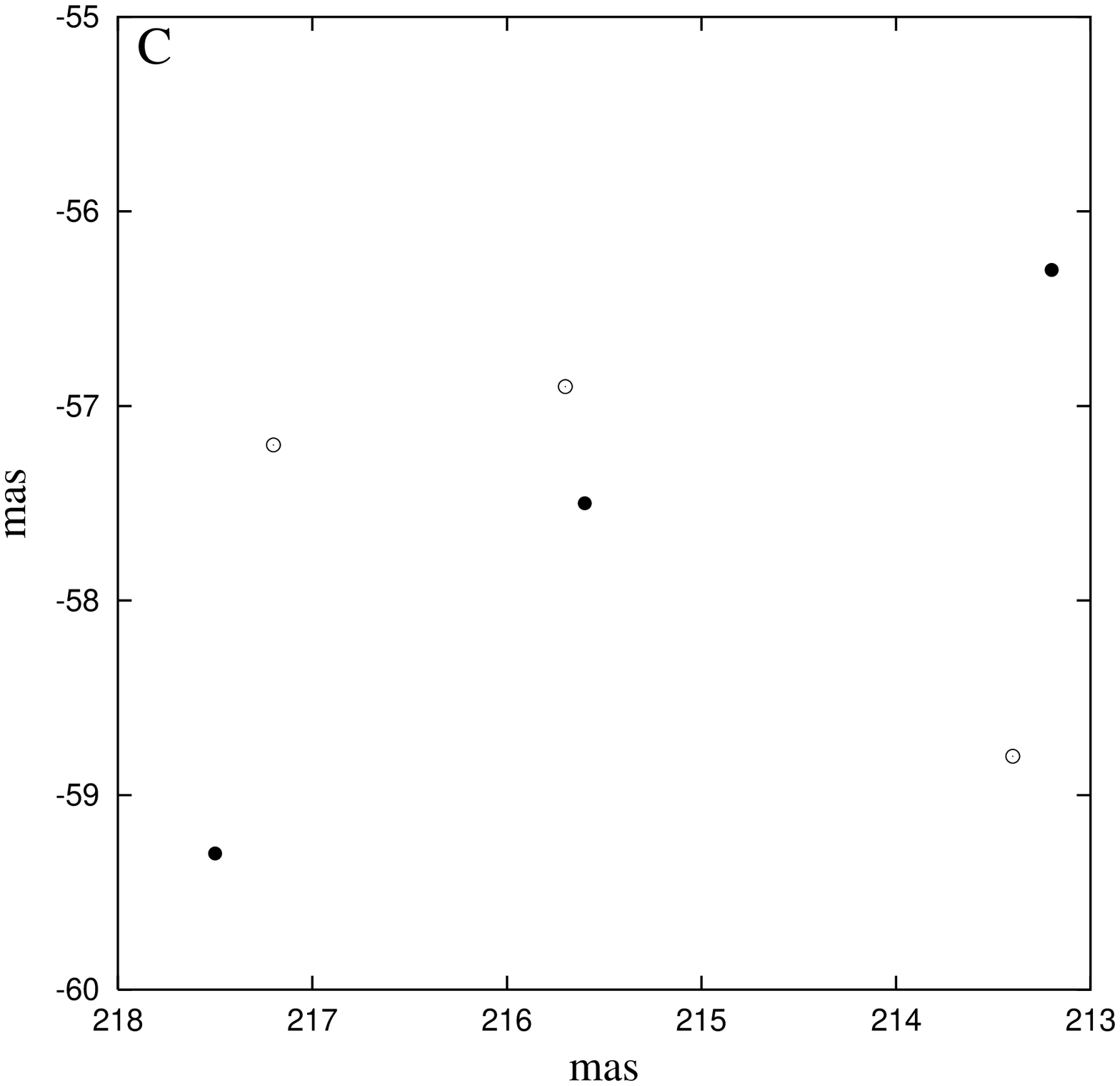}
\includegraphics[scale=0.33,bb=50 50 554 554,angle=-90]{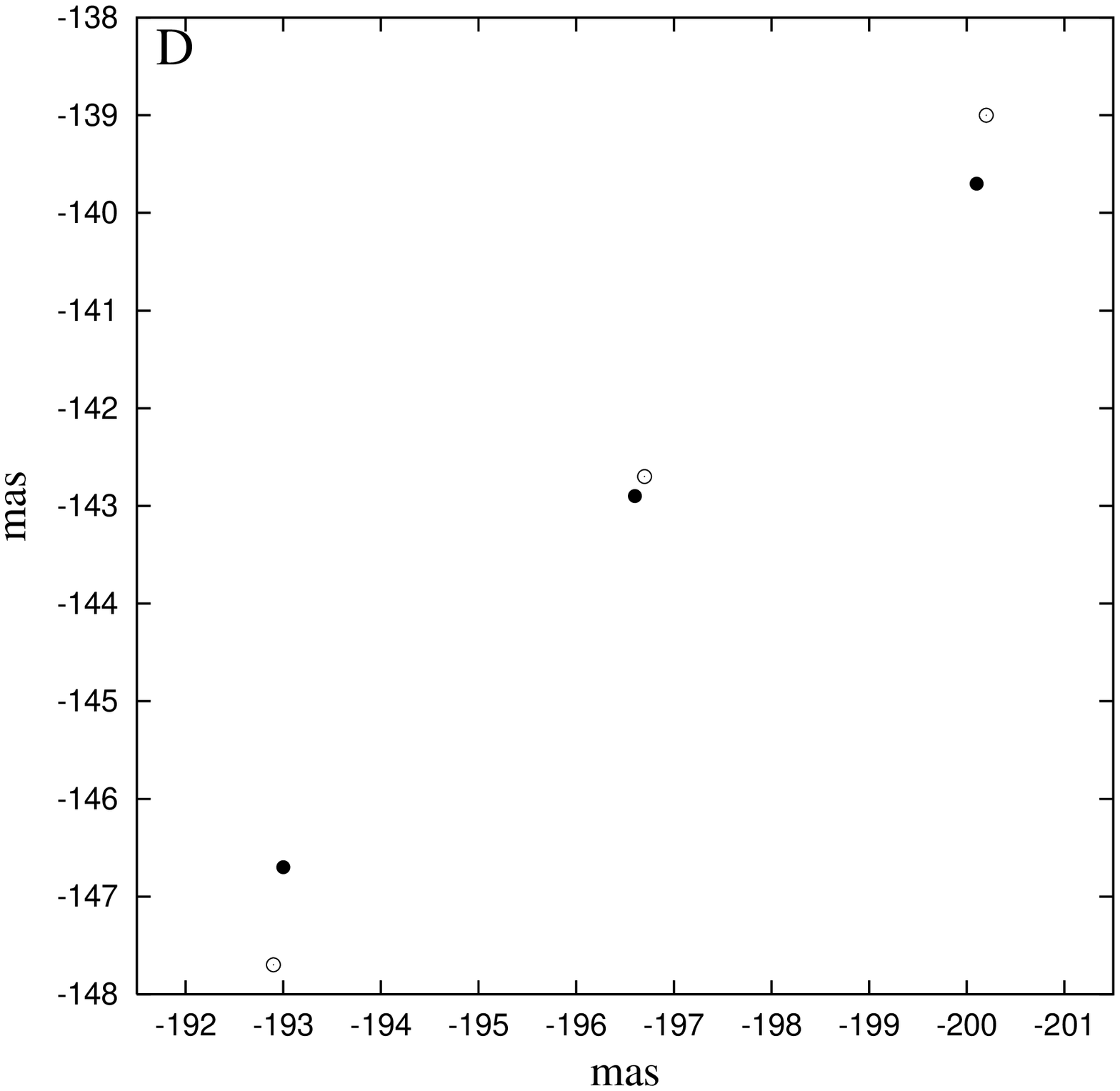}
\caption{Observed positions of image sub-components (filled circles)
  and derived model positions (open circles) for Model 2 using the {\sc
  glint} lens modelling software. The image B positional constraints
  have been estimated from the VLBI maps.}
\label{model2}
\end{center}
\end{figure*}

Increasing the errors on images A, C and D by an order of magnitude
to 1~mas (Model~2) produces a better fit to image B and the expected
improvement in the goodness-of-fit ($\chi^2 \sim 25$
and $\overline{\chi}^2 \sim 2$). The model positions are plotted in
Fig.~\ref{model2} and the lens galaxy parameters are again shown in 
Table~\ref{lensmodtab}. The model now places image B in the correct
area of the image plane and the offsets between the observations and
model lie between only 3 and 6~mas. These act to make the modelled jet
too long, by about a factor of two. The fit to image C, conversely, has
deteriorated. The offsets of the end sub-components have increased, to
$>$2~mas, thus accentuating a bend in the jet that was evident in 
Fig.~\ref{model1}. Images A and D again show no signs of any
significant discrepancy.

These two models seem to highlight a degeneracy between
the ability to fit either image B or image C, but not both. For
example, we have also investigated the case when the image B positions
are completely unconstrained in the modelfit i.e. their errors are made
very large. In this latter case we obtain $\chi^2 \sim 50$ which, with
now only 5 degrees of freedom, gives a reduced chi-squared of $\sim$10.
Excellent fits are found for images A and D, the observed positions
being recovered to within the formal uncertainties. The modelled
positions for image B, though, are highly discrepant, lying
about 100~mas away from the observed position, a consequence of this
image being completely unconstrained in the fit. Image C though is now
fit relatively well, the offsets between modelled and observed
positions being less than 0.5~mas for both C1 and C3.

Further simulations demonstrate the malleability of the model when
various combinations of position errors are used as constraints. For
example, we have found it possible to produce a progressively poorer
fit to image A by reducing the errors on image C incrementally from
1~mas to 0.1~mas, whilst leaving the errors on A and D at 1~mas. This is
accompanied by the ellipticity of the SIE increasing to $\sim$0.9. We
also point out that the usual degeneracy between SIE ellipticity and
external shear is present. This makes it very likely that different
executions of a given modelling package (as well as different packages
that use different optimisation routines) will not converge to the same
minimum.

However, the clear conclusion is that no smooth model can be obtained
that fits all the astrometric constraints; it is not difficult to get a
reasonable fit to images A and D but fitting both C and D
simultaneously is not possible.

\begin{table}
\begin{center}
\caption{Parameters for the SIE+external shear lens model. Listed are
  the galaxy Einstein 
  radius ($\theta_E$), position ($x, y$), magnitude and position angle
  of ellipticity ($\epsilon, \theta_{\epsilon}$) and magnitude and
  position angle of external shear ($\gamma, \theta_{\gamma}$).
  Position angles are measured North through East and the galaxy
  position is offset from the VLBI phase centre ($x$ increases to the
  East as with Right Ascension). For a description of the models see
  the text.}
\begin{tabular}{lll} \hline
Parameter & Model 1 & Model 2 \\ \hline
$\theta_E$ & 229~mas & 237~mas \\
$x, y$ & $-$81~mas, $-$4~mas & $-$79~mas, $+$4~mas\\
$\epsilon, \theta_{\epsilon}$  & 0.42, $-$51$^{\circ}$ & 0.41,
$-$36$^{\circ}$ \\
$\gamma, \theta_{\gamma}$ & 0.26, 29$^{\circ}$ & 0.22, 37$^{\circ}$ \\
\end{tabular}
\label{lensmodtab}
\end{center}
\end{table}

\subsection{\citeauthor{evans03} method}

Recently, \citet{evans03} have introduced a method for lens modelling
that models the lensing mass as a sum of Fourier components, the
advantage of this being that the azimuthal profile is allowed to be
completely arbitrary (whilst at the same time the galaxy is constrained
to be isothermal and scale-free). This allows galaxies to be fit that
are, for example, boxy or which contain bar-like structures. They
have used their method to demonstrate that the anomalous flux ratios
seen in some lens systems can, in some cases, be fit with a perfectly
smooth galaxy, thus negating the need for CDM substructure to be
invoked.

We have implemented our own version of the \citeauthor{evans03} method
(with which we can reproduce the authors' results) and applied it to
B0128+437. Initially we have only modelled a single source and find
that it is relatively easy to fit a smooth galaxy, as should be
expected given that the SIE+external shear model also works well in
this case. However, when we increase the number of sources, first to
two and then to three, we find it increasingly difficult to find a fit
where the galaxy has smooth isophotes.

The optimisation problem is, though, very complicated. The matrix that
needs to be inverted (through singular-value decomposition) is very
large for three sources ($33 \times 33$) and even with our use of a
simulated annealing optimisation algorithm, it is not clear how
successful we are in locating the global minimum. An additional
complication is that singular values which become very small need to be
discarded. The choice of the threshold below which these are removed is
crucial as this removes information and results in a
smoother galaxy. For example, we find that a threshold of 10$^{-5}$ of
the largest singular value (with two sources modelled) produces similar
results to a threshold of 10$^{-4}$ (for three sources).

\begin{figure*}
\begin{center}
\includegraphics[scale=0.45]{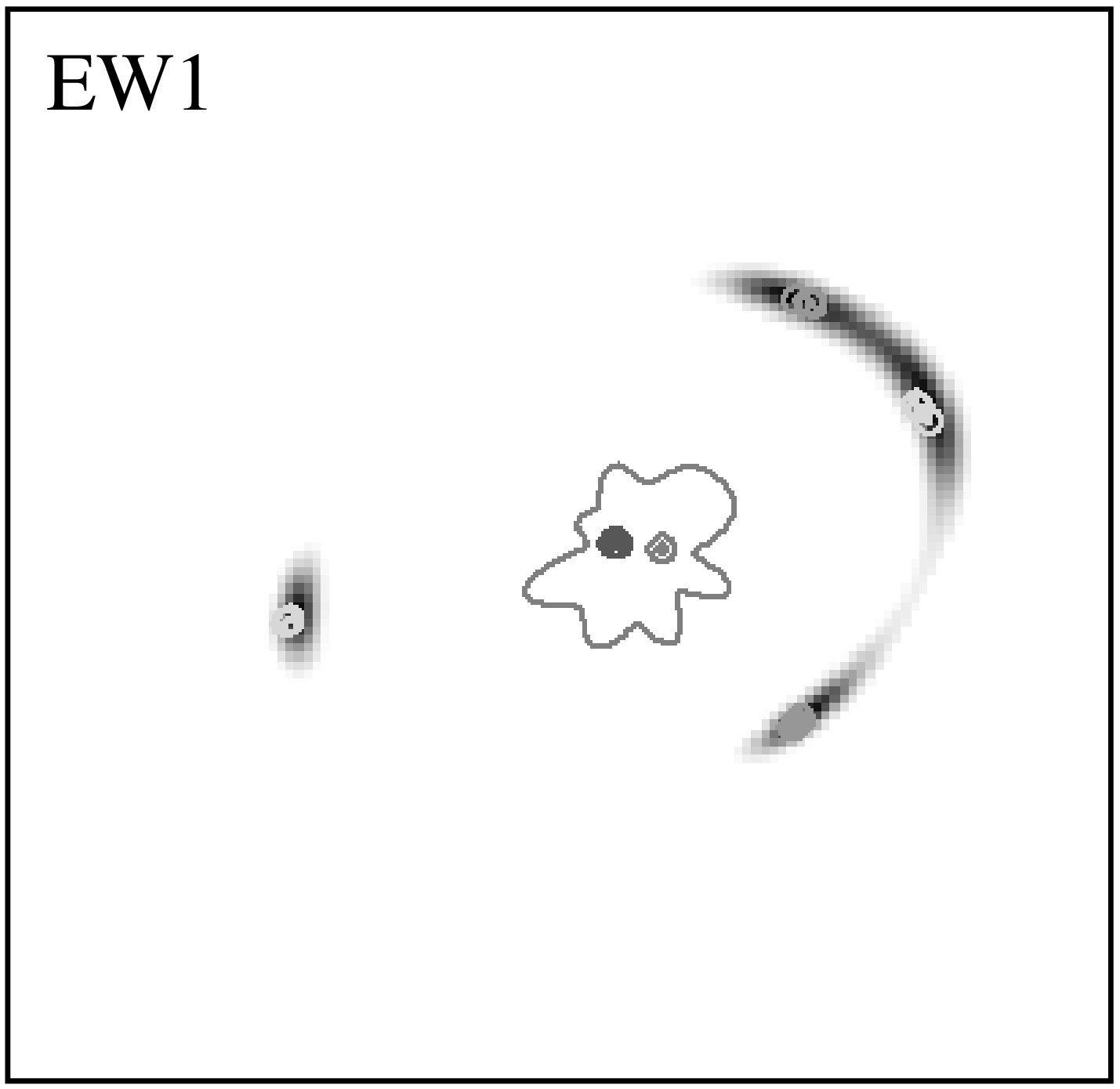}
\includegraphics[scale=0.45]{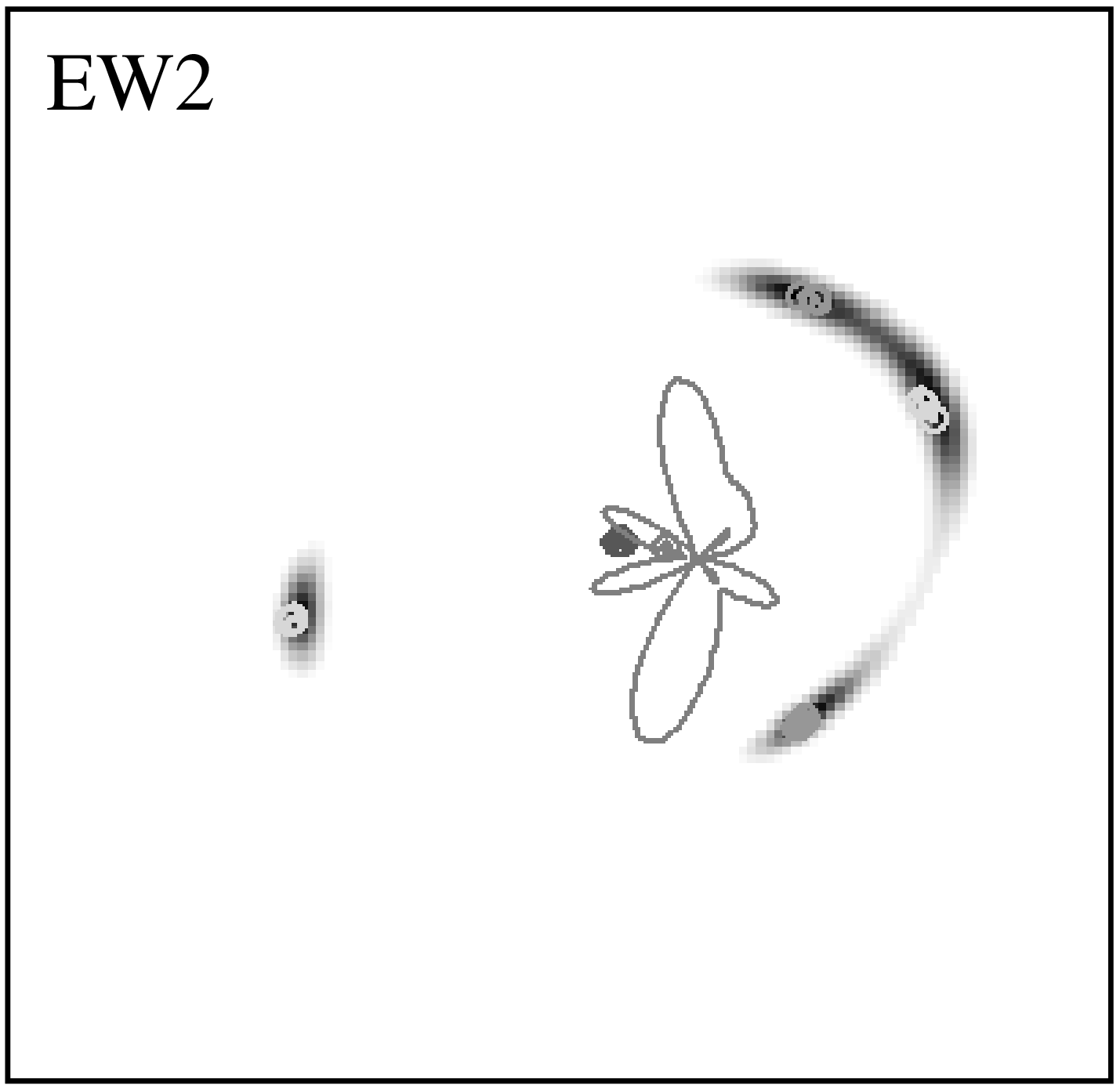}
\caption{Results of applying the \citeauthor{evans03} method to all
  three sources of
  B0128+437. The model on the left (EW1) uses 1~mas errors on the
  A, C and D sub-component positions whilst these are decreased to
  0.1~mas in the model on the right (EW2). In both instances the errors
  on the image B positions are 5~mas and the singular-value threshold
  10$^{-4}$. The thin grey lines represent equidensity contours of the
  lensing mass. The remainder of the plots show the observed (circles)
  and modelled (greyscales) image positions for an SIE+external shear
  model for scale. The circles close to the centre of the equidensity
  contours are the source and galaxy core (lighter of the two)
  positions.}
\label{ewfig}
\end{center}
\end{figure*}

Bearing the above caveats in mind, we find results that are consistent
with those found in the SIE+external shear modelling, using all three
sources. With 1~mas errors on the image A, C and D sub-components and
5~mas on B (EW1) we can produce a much smoother galaxy than we can
using the nominal astrometric errors of 0.1~mas (EW2), for the same
threshold ($10^{-4}$). In Fig.~\ref{ewfig} we show the equidensity
contours for these two models. The galaxy on the left (large errors) is
clearly much smoother than that on the right which, in common with the
SIE method, is consistent with the galaxy containing substructure.

\section{Discussion}

Based on the initial work presented by \citet{phillips00} it appeared
as though CLASS B0128+437 was a fairly ordinary lens system that could
be simply and satisfactorily modelled. The observations and modelling
presented here instead show that this system is problematic in at least
two respects.
\begin{enumerate}
\item The strange appearance of image B, especially the absence of the
  sub-components seen in the other images.
\item The difficulty in fitting a mass model to the observed positions
  of the image sub-components.
\end{enumerate}
We now go on to discuss the origin of these as well as considering the
flux ratios of the lensed images and the nature of the lensed source.

\subsection{Scatter-broadening of image B}
\label{discussscattering}

The most obvious and intriguing feature of the VLBI images of B0128+437
is that there is little sign in image B of the three discrete
sub-components that are so obvious in the other three images. However,
examining the maps very carefully, it seems that B2 and B3 are in fact
detected, albeit weakly. We identify these as the two weak patches of
emission labelled in the uniformly-weighted 5-GHz map of
Fig.~\ref{mapscuni}.
This shows that the brightest end of the jet is actually made up of two
discrete components separated by about $\sim$5~mas, similar to the
separation between sub-components A2 and A3. Their location also places
them at the correct end of the jet (the end nearest image A), the
natural identification for the very faint emission seen at 8.4~GHz at
the opposite end of the jet then being that this is probably the
remnants of B1. We also note that the sub-component that we identify as
B2 has a higher surface brightness than B3, again in common with the
equivalent sub-components in A and D. The signal to noise is, however,
low and the distortion of image B renders this identification far from
definite.

But what is causing image B to be so faint? The most likely 
explanation we believe is that this image is scatter-broadened in a
turbulent ionised medium. Contrary to the basic principles of
gravitational lensing, surface brightness is clearly not conserved in
image B, unlike in the other images (seen most clearly at
8.4~GHz). Scatter-broadening would naturally account for this as the
flux contained within an image would be spread out over a larger area
and thus appear fainter. The most convincing evidence for this effect
in image B is its very different appearance in the uniform and natural 
maps, being very much stronger with natural weighting, whilst the other
images look relatively similar under both weighting schemes. For
example, whilst between 60 and 75~per~cent of the naturally-weighted
total flux density is detected in the uniform maps of images A, C and
D (as measured in an aperture around the image using the {\sc aips}
task {\sc tvstat}) this
drops to only 30~per~cent for image B (Table~\ref{tvstat}). This
indicates that the angular scale of image B is significantly larger
than that of the other images, despite having a lower or comparable
flux density. There is also no sign of the gap that exists
approximately midway between A1 and A2 in image B.

\begin{table}
\begin{center}
\caption{5-GHz flux density of each image as measured from the VLBA
  maps, for both uniform and natural weighting.}
\begin{tabular}{lcccc} \hline
 & A & B & C & D \\ \hline
$S_{5,\mathrm{uni}}$ (mJy) & 10.6 & 2.8 & 4.8 & 4.8 \\
$S_{5,\mathrm{nat}}$ (mJy) & 14.0 & 9.1 & 7.9 & 8.3 \\
$S_{5,\mathrm{uni/nat}}$ (mJy) & 0.76 & 0.31 & 0.61 & 0.58 \\ \hline
\end{tabular}
\label{tvstat}
\end{center}
\end{table}

A number of other features of image B support the scattering
hypothesis. Our multi-frequency maps show that the surface brightness
of image B is not only lower than the other images, but that it has a
different spatial and frequency dependence from the other images. At
5~GHz the surface brightness of image B drops smoothly from its peak at
the south-western area where B2 and B3 are located before fading away
entirely around the location of B1. At 2.3~GHz, however, the reduction
in surface brightness is much more pronounced, dropping very suddenly
at around $-$200, 210~mas. It is clear from a comparison between
Figs~\ref{mapsc} and \ref{mapss} that the area in between the two
dashed circles is brighter at 5~GHz than at 2.3~GHz, despite the
surface brightness sensitivities of the two arrays being similar and
the source being brighter at lower frequencies (above $\sim$1~GHz). 
As scattering produced by a screen with uniform statistical properties
over regions large in angular size compared to the source increases
angular sizes proportional to the wavelength squared, the observed
frequency dependence is in the correct sense. The spatial dependence is
seen best at 8.4~GHz where similarly very weak emission is detected
from all three sub-components despite the fact that B1 should be much
brighter.

If scatter-broadening is the explanation for the anomalous appearance
of image B, there must be a cloud of clumpy ionised gas along the line
of sight to image B and the strength of the scattering must increase
towards the north-eastern end of the jet. The spatial dependence of the
scattering within image B emphasises the fact that the scattering
within this lens system is non-uniform i.e. image B alone is obviously
affected. The most likely location of the scattering medium therefore
is the lensing galaxy as this is the obvious location along a line of
sight between the radio source host galaxy and our Galaxy where a large
amount of ionised gas could reside and affect the images to different
degrees. The small separation of the images predicts a relatively low
mass for the lensing galaxy which may therefore be a spiral galaxy and
rich in gas and dust. Other lens systems where images are believed to
be scatter-broadened include PMN~J0134-0931 \citep{winn03b},
JVAS~B0218+357 \citep{biggs03}, PKS~1830-211 \citep{jones96,guirado99}
and CLASS~B1933+503 \citep{marlow99}. 

As a test for scattering elsewhere in B0128+437 we can examine the
model fits to A1 and D1 for which we are able to measure sizes at two
frequencies. We do not consider C1 as the small separation of the
sub-components in this image renders the model fits, especially their
sizes, unreliable. Assuming that the angular size of the images (as
measured by the geometrical mean of the major and minor axes, 
$\theta_{\mathrm{g}}$) scales as $\theta_{\mathrm{g}} \propto
\lambda^{\alpha}$, we measure $\alpha = 1.6\pm0.2$ for A1 and 
$\alpha = 2.5\pm0.4$ for D1. These increases are greater than the
linear dependence on frequency usually expected for an optically-thin
synchrotron source and suggest that these images may also be affected
by scattering. The haloes surrounding images C and D may also be a
consequence of scatter-broadening as similar low surface brightness
regions either side of the jet are not seen in the other images. We
also note that although C2 is rather faint in Fig.~\ref{mapscuni}
(uniformly-weighted data), this is apparently not the case when natural
weighting is used i.e. this sub-component, in common with image B, may
be being scatter-broadened. However, the low resolution
makes it difficult to be sure of this. (This flux ratio anomaly is also
considered in the context of galaxy substructure in
Section~\ref{discusssub}.)

We have considered other explanations for the anomalous appearance of
image B, such as a lensing distortion due to compact structures within
the lensing galaxy and free-free absorption, but these cannot explain
all the effects observed in this image including its larger size
relative to the other images as well its frequency-dependent reduction
in surface brightness.

\subsection{Lens substructure}
\label{discusssub}

The SIE lens modelling that we have undertaken has demonstrated that it
is not possible to reproduce the observed model positions of the VLBI
sub-components in B0128+437 unless we assume that the nominal
error bars on the sub-component positions are large underestimates of
the true uncertainties. Using the nominal errors produces a poor fit to
the data, the most striking aspect of this being that image B is offset
by tens of mas from its observed position. While there is of course
some uncertainty in the positions of the image B sub-components, it is
certainly not greater than 5~mas, and less if our identification of B2
and B3 in Fig.~\ref{mapscuni} is correct. The fact that we have used
error bars on the image B positions which are 50 times larger than
those on the other images explains why this image is so poorly
constrained during the optimisation.

Most striking about the model is the degeneracy we find between the
model's ability to fit the positions of B and C. The best fit for image
C is found when image B is completely unconstrained (its modelled
position ending up very close to that of image A) whilst the better the
agreement with the observed position of image B, the poorer the fit to
image C. Given that we must insist that the model produces positions
for image B that are consistent with the observations to at least
several mas, the coupling we find between the positions of images B and
C means that we must accept that the model cannot reproduce the
positions of the sub-components in image C. 

As we can not fit the observed image positions with a smooth
parameterisation of the mass in the lensing galaxy (a conclusion which
is also suggested by the results of the \citeauthor{evans03} method)
then there must be additional unmodelled mass in the deflector,
i.e. substructure. 
Hierarchical models of large-scale structure formation such as
Cold Dark Matter (CDM) predict that galactic haloes should contain a
population of satellites with masses that extend down to at least
10$^7$M$_{\sun}$, the current resolution of the best simulations
\citep[e.g.][]{moore99,delucia04}. Our own galaxy contains an order of
magnitude less satellites than predicted and many authors have recently
considered gravitational lensing as a way of implying the presence of
these objects in lensing galaxies, mainly through the detection of
anomalous flux ratios \citep[e.g][]{dalal02,metcalf02a}, but also
through image position shifts and distortions
\citep[e.g.][]{wambsganss92,metcalf01,metcalf02b}.

We believe that the need to increase the uncertainties on the
sub-component positions (by a factor of ten) to 1~mas in order to gain
an acceptable fit is evidence that some form of substructure is present
in the lens galaxy of B0128+437. The modelled position shifts of
several mas imply substructure of at least $\sim$10$^6$M$_{\sun}$. If a
perturbing mass was located close to image C, this could also affect
the flux ratios of the image C sub-components, perhaps explaining why
the central sub-component, C2, is so faint in Fig.~\ref{mapscuni} (but
see Section~\ref{discussscattering} for an alternative hypothesis).

Conclusions based on the lens mass modelling should be strengthened
with improved optical/infrared images of the lens system with which the
properties of the lensing mass (positions, shape, etc) could be
measured and used as model constraints. This may be possible with
approved $HST$ NICMOS observations that should take place in 2004.

\subsection{Flux density ratios}

\begin{table}
\begin{center}
\caption{Flux density ratios for various image combinations at four
  frequencies (X: 8.4~GHz, C: 5~GHz, S: 2.3~GHz, L: 1.4~GHz).}
\begin{tabular}{lcccc} \hline
  & VLA (X) & MERLIN (C) & VLBA (S) & MERLIN (L) \\ \hline
$S_{\mathrm{B/A}}$ & -    & 0.56 & 0.49 & -    \\
$S_{\mathrm{C/A}}$ & -    & 0.49 & 0.34 & -    \\
$S_{\mathrm{D/A}}$ & -    & 0.47 & 0.47 & -    \\
$S_{\mathrm{C/D}}$ & 0.67 & 1.04 & 0.72 & 0.75 \\ \hline
\end{tabular}
\label{fluxratio}
\end{center}
\end{table}

The radio data do not allow the production of spectra for each image
across a very wide range of frequencies. The only data that cleanly
separate each image, other than those obtained with VLBI, are the
MERLIN 5-GHz data although MERLIN at 1.4~GHz and the VLA at 8.4~GHz do
resolve images C and D. We consider the flux densities measured from
the VLBI data to be less accurate than those from the VLA and MERLIN
data, but the 2.3~GHz image shown in Fig.~\ref{optical} (where the
images are nearly at their brightest and where a ($u,v$) taper has been
applied thus increasing the sensitivity to low surface brightness
emission) should be fairly reliable.

We have estimated flux densities at each frequency
by model fitting to the visibilities, except at 2.3-GHz. In this case we
use a variety of methods including summing the CLEAN components,
measuring the flux contained within an aperture as well as summing all
pixels that have a surface brightness greater than 6$\sigma$ of the rms
map noise. All methods gave broadly consistent results and we have
taken their average as the best estimator at this frequency. We only
consider the flux ratios as these are not subject to errors in the
absolute flux density scale and show these in Table~\ref{fluxratio}.

Given $\la$5~per~cent errors on the flux density ratios, the most
surprising thing about the numbers in Table~\ref{fluxratio} is the
flux density of image C relative to image D. At three of the four
frequencies (1.4, 2.3 and 8.4~GHz) C is significantly fainter than D
whilst at 5~GHz the flux densities of each are the same to within the
errors. The 5-GHz data would not appear to be in error as the same flux
density ratios were measured from 41 epochs of data that were
observed as part of a project to look for radio microlensing in a 
sample of JVAS/CLASS lens systems \citetext{\citealt{koopmans03a};
Biggs et al., in preparation}. Similarly, the  
fact that a ratio of $\sim$0.7 is measured at the other three
frequencies strongly argues that these measurements are also robust and
that the flux ratio anomaly is real. The MERLIN monitoring data also
showed that the flux density of each image was constant over the
eight-month monitoring period which rules out intrinsic source
variability as the cause and demonstrates that the effect is
long-lived. Interpretation of this effect would be greatly aided by
additional observations that would allow the flux density of each image
to be measured accurately over a range of frequencies. This could be
done with the VLA at 15 and 22~GHz, in conjunction with the VLBA Pie
Town antenna at 8.4 and 5~GHz. 

\subsection{The lensed radio source}

The radio source in B0128+437 is a member of the GPS class as its
spectrum peaks at around 1~GHz. This allowed \citet{phillips00} to
estimate the maximum angular separation that would be observed in
VLBI observations based on a correlation of this between the turnover
frequency and the source flux density \citep{snellen00}. Our
observations confirm their prediction of substantial resolved structure
in this system.

We are also able to measure the spectral indices of the individual
sub-components between 5 and 8.4~GHz. This is done most easily for
sub-component 1, the only particularly bright one in each image at
8.4~GHz, where we measure $\alpha \sim -0.16$ for A1 and D1, where
$S_{\nu} \propto \nu^{\alpha}$. Model fitting to the 8.4-GHz ($u,v$)
data gives a flux density for A2, the brightest sub-component 2 image,
of $\sim$1.6~mJy which corresponds to a much steeper spectrum of
$\alpha \sim -1.27$. The third sub-component also has a steep spectrum,
but this is difficult to measure due to its weakness in all images at 
8.4~GHz.

The fact that one of the outer sub-components has a flat spectrum
suggests that this is the core and that the source is a classical
core-jet. However, at 5~GHz (the only frequency where reliable sizes
can be measured for all three sub-components) the most compact
sub-component is the central one. Furthermore, emission is detected on
either side of the flat-spectrum sub-component at both 5 and 2.3~GHz.
Both of these observations are unusual as it is more commonly the
flat-spectrum core component that is the most compact and core-jet
sources, which are believed to point more or less towards the observer,
should not display counterjet emission as this should be greatly
reduced in brightness due to Doppler de-boosting.

Also unusual is the presence of low-brightness extended emission
in the 2.3~GHz maps. This would not be expected given that the source
spectrum peaks at 1~GHz (the total flux density at 325~MHz is only
$\sim$15~mJy \citep{phillips00}). B0128+437 is, in this respect, similar
to another lens system, JVAS~B2114+022, where two of the components (which
may or may not be lensed) are very diffuse, despite having spectra that
turnover above 1~GHz \citep{augusto01}. It may be that the
turnover in the spectrum is not intrinsic to the source, but a
consequence of a propagation effect such as free-free absorption. Another
possibility is that the spectrum is intrinsic and that the low-brightness
extended emission results from scatter-broadening, as has already been
suggested for image B.

\section{Summary and Conclusions}

We have presented VLBI observations of the gravitational lens system
CLASS B0128+437 that reveal extensive structure within each of the four
images. These data illustrate the role that high-resolution VLBI
observations play in the study of lens systems, such as providing a
large number of model constraints and identifying inconsistencies 
between the observations and the lens model. Interesting astrophysical
conclusions regarding the lens galaxy can be drawn that would
otherwise be missed using data from smaller arrays such as MERLIN and
the VLA. Our observations demonstrate that despite the large number of
model constraints made available by the VLBI imaging, we find it
difficult to get smooth mass models to fit the observed image
positions, signifying additional unmodelled structure in the
deflector. This may be another instance of lensing providing evidence
to support the existence of mass substructure in CDM haloes. Improved
infrared $HST$ observations will be taken which may improve our
understanding of the deflecting mass, leading to more rigorous
modelling.

The most striking feature of our new VLBI maps is the anomalous
appearance of image B compared to the other images. The most likely
explanation for this is that this image is being scatter-broadened in
the ISM of the lensing galaxy, a process which may also be affecting
the other images to a lesser extent. An important consequence of this
effect concerns the 
identification of lens systems in CLASS. The CLASS methodology was to
use VLBA 5-GHz observations, once candidate systems had been identified
with the VLA and MERLIN, to see whether any could be rejected on the
basis of surface brightness arguments i.e. weaker components more
extended, or through inconsistent structures in the different
images. In a two-image system particularly, were one of the images to
be distorted in a manner similar to that seen in B0128+437, that system
may have been rejected. We are currently re-observing 13 CLASS
two-image lens candidates at 15~GHz with the VLBA in order to check
whether their initial rejection was sound.

\section*{Acknowledgments}

MAN, JPM, PMP and TY acknowledge the receipt of PPARC studentships. The
VLBA is operated by the National Radio Astronomy Observatory which is a
facility of the NSF operated under cooperative agreement by Associated
Universities, Inc. This work included observations with the 100-m
telescope of the MPIfR (Max-Planck-Institut f\"{u}r Radioastronomie) at
Effelsberg. This work also included observations made with the NASA/ESA
$Hubble Space Telescope$, obtained at the Space Telescope Science
Institute, which is operated by AURA, Inc., under NASA contract NAS
5-26555. UKIRT is operated by the Joint Astronomy Centre on behalf of
PPARC. This publication makes use of data products from the Two Micron
All Sky Survey, which is a joint project of the University of
Massachusetts and the Infrared Processing and Analysis
Center/California Institute of Technology, funded by the National
Aeronautics and Space Administration and the National Science
Foundation. We thank the referee, David Rusin, for his comments.

\bibliographystyle{mnras}
\bibliography{0128}

\end{document}